%% file: range-locks.tex
\documentclass[sigplan,10pt]{acmart}

\renewcommand\footnotetextcopyrightpermission[1]{}

\usepackage{microtype}
\usepackage{listings}

\usepackage{graphicx}
\usepackage[skip=5pt]{caption}
\usepackage{subfig}
\usepackage[export]{adjustbox}
\usepackage{url}

\usepackage[small,compact]{titlesec}
\usepackage{xspace}

\usepackage[colorinlistoftodos]{todonotes}

\input{defs.tex}

\definecolor{codegreen}{rgb}{0,0.6,0}
\definecolor{codegray}{rgb}{0.5,0.5,0.5}
\definecolor{codepurple}{rgb}{0.58,0,0.82}
\definecolor{backcolour}{rgb}{0.95,0.95,0.92}

\lstdefinestyle{mystyle}{
  backgroundcolor=\color{backcolour},   commentstyle=\color{codegreen},
  keywordstyle=\color{magenta},
  numberstyle=\tiny\color{codegray},
  stringstyle=\color{codepurple},
  basicstyle=\footnotesize,
  breakatwhitespace=false,
  breaklines=true,
  captionpos=b,
  keepspaces=true,
  numbers=left,
  numbersep=5pt,
  showspaces=false,
  showstringspaces=false,
  showtabs=false,
  tabsize=2
}

\def\ContinueLineNumber{\lstset{firstnumber=last}}
\def\StartLineAt#1{\lstset{firstnumber=#1}}

\makeatletter
\lst@Key{countblanklines}{true}[t]%
    {\lstKV@SetIf{#1}\lst@ifcountblanklines}
    
\lst@AddToHook{OnEmptyLine}{%
    \lst@ifnumberblanklines\else%
       \lst@ifcountblanklines\else%
         \advance\c@lstnumber-\@ne\relax%
       \fi%
    \fi}
\makeatother

\lstdefinestyle{numbers}
{numbers=left, stepnumber=1, numberstyle=\tiny, numbersep=10pt}
\lstdefinestyle{nonumbers}
{numbers=none}

\newtheorem{invariant}{Invariant}
\usepackage{ifthen}
\usepackage{xcolor}

\newboolean{showcomments}
\setboolean{showcomments}{true}
\ifthenelse{\boolean{showcomments}}
{ \newcommand{\mynote}[3]{
    \fbox{\bfseries\sffamily\scriptsize#1}
    {\small\textsf{\emph{\color{#3}{#2}}}}}}
{ \newcommand{\mynote}[3]{}}

\usepackage{fancyhdr}
\usepackage{tcolorbox}
\fancypagestyle{firstpage}{
\fancyhead{
    \vspace{-35pt}
    \begin{tikzpicture}
        \node[align=center] () at (0,0) {
            \begin{tcolorbox}[colback=yellow!40,
                              colframe=white,
                              width=\textwidth,
                              boxrule=0mm,
                              sharp corners]
                    \centering
                    This is an extended version of the article published in the proceedings of the 2020 15th European Conference on Computer Systems (EuroSys'20), which is available online at \url{https://doi.org/10.1145/3342195.3387533}.\\
            \end{tcolorbox}
        };
    \end{tikzpicture}
}
\fancyfoot{}
\cfoot{\thepage}
}

\title[Scalable Range Locks for Scalable Address Spaces and Beyond]{Scalable Range Locks for\\ Scalable Address Spaces and Beyond}

\fancypagestyle{normal}{
\lhead{\shorttitle}
\rhead{\shortauthors}
\fancyfoot{}
\cfoot{\thepage}
}

\begin{document}

\author{Alex Kogan}
\affiliation{
  \institution{Oracle Labs}
  \city{Burlington}
  \state{MA}
  \country{USA}
}
 \email{alex.kogan@oracle.com}

\author{Dave Dice}
\affiliation{
  \institution{Oracle Labs}
  \city{Burlington}
  \state{MA}
  \country{USA}
}
 \email{dave.dice@oracle.com}

\author{Shady Issa}
\authornote{Work was done while the author was an intern at Oracle Labs.}
\affiliation{
  \institution{U. Lisboa \& INESC-ID}
  \city{Lisbon}
  \country{Portugal}
}
\email{shadi.issa@tecnico.ulisboa.pt}


\begin{abstract}
Range locks are a synchronization construct designed to provide concurrent access 
to multiple threads (or processes) to disjoint parts of a shared resource.
Originally conceived in the file system context, range locks are gaining 
increasing interest in the Linux kernel community seeking to alleviate bottlenecks in the 
virtual memory management subsystem.
The existing implementation of range locks in the kernel, however, uses an internal spin lock 
to protect the underlying tree structure that keeps track of acquired and requested ranges.
This spin lock becomes a point of contention on its own when the range lock is frequently acquired.
Furthermore, where and exactly how specific (refined) ranges can be locked remains an open question.

In this paper, we make two independent, but related contributions.
First, we propose an alternative approach for building range locks based on linked lists.
The lists are easy to maintain in a lock-less fashion, and in fact, our range locks do not use any internal locks
in the common case.
Second, we show how the range of the lock can be refined in the \code{mprotect} operation through a speculative mechanism.
This refinement, in turn, allows concurrent execution of \code{mprotect} operations on non-overlapping memory regions.
We implement our new algorithms and demonstrate their effectiveness in user-space and kernel-space,
achieving up to 9$\times$ speedup compared to the stock version of the Linux kernel.
Beyond the virtual memory management subsystem, we discuss other applications of range locks in parallel software.
As a concrete example, we show how range locks can be used to facilitate the design of scalable concurrent data structures, such as skip lists.



\end{abstract}


\settopmatter{printacmref=false}


\maketitle
\thispagestyle{firstpage}

\section{Introduction}
Range locks are a synchronization construct designed to provide concurrent access 
to multiple threads (or processes) to disjoint parts of a shared resource.
Originally, range locks were conceived in the context of file systems~\cite{unix-manual}, 
to address scenarios in which multiple writers would want to write into different parts of the same file.
A conventional approach of using a single file lock to mediate the access
among those writers creates a synchronization bottleneck.
Range locks, however, allow each writer to specify (i.e., lock) the part of the file it is going to update, 
thus allowing serialization between writers accessing the same part of the file, 
but parallel access for writers working on different parts.

In recent years, there has been a surge of interest in range locks in a different context.
Specifically, the Linux kernel community considers using range locks to address contention on 
\texttt{mmap\_sem}~\cite{Duf17}, which is ``one of the most intractable contention points in the memory-management subsystem''~\cite{Cor17}.
\texttt{mmap\_sem} is a \rw semaphore protecting the access to the virtual memory area (VMA) structures.
VMA represents a distinct and contiguous region in the virtual address space of an application;
all VMA structures are organized as a red-black tree (\code{mm\_rb})~\cite{CKZ12}.
The \texttt{mmap\_sem} semaphore is acquired by any virtual memory-related operation, 
such as mapping, unmapping and mprotecting memory regions, and handling page fault interrupts.
As a result, for data intensive applications that operate on chunks of dynamically allocated memory,
the contention on the semaphore becomes a significant bottleneck~\cite{Cor17, CKZ12, DK19b}.

The existing implementation of range locks in the Linux kernel is relatively straightforward.
It uses a range tree (based on red-black trees) protected by a spin lock~\cite{Kar13}.
Given that every acquisition and release of the range lock, for any range,
results in the acquisition and release of that spin lock, the latter can easily become a bottleneck on its 
own under heavy use regardless of the contention on actual ranges.
Note that even non-overlapping ranges and/or ranges acquired for read have to synchronize 
using that same spin lock. 
We expand on the implementation of existing range locks in the kernel and its shortcomings in \secref{existing-range-locks}.

Even when putting the issues in the existing range lock implementation aside,
exploiting the potential parallelism when using range locks to protect the access
to VMA structures in the Linux kernel is far from trivial.
The key challenge is that addresses presented to virtual memory (VM) operations 
(singular addresses arising from page fault handling or ranges associated with APIs such as \code{mprotect}) 
do not necessarily fall on VMA boundaries.
Thus, the enclosing range of the VM space that needs to be protected is not known in advance of walking the \code{mm\_rb} tree. 
Therefore, simply applying a VM operation under the lock acquired for the range of that operation does not work.
As an intuitive example, consider two \code{mprotect} operations on different (non-overlapping) memory ranges.
If those operations acquire the range lock only on those (non-overlapping)  ranges, they may race with each other on 
updates to the VMA metadata if they end up operating on the same VMA.
Furthermore, regardless of whether two \code{mprotect}s operate on the same VMA, 
if one of them rotates the \code{mm\_rb} tree, the other one may read an inconsistent state 
while traversing the tree in parallel.
All these issues might be the reason that in the kernel patch that replaces \code{mmap\_sem} with a range lock,
the latter is always acquired for the full range\footnote{The range lock API 
includes calls to acquire the lock for a specific range (e.g., [$10$..$25$]) as well as a special call to acquire the lock for the 
entire (full) range (i.e., [$0$..$2^{64}-1$]).}
~\cite{Bue18}, exploiting no potential parallelism that range locks can provide\footnote{The author of the patch notes that 
"while there is no improvement of concurrency perse, these changes aim at adding the machinery to permit this in the future."
We are not aware of any follow-up work that does that.}.

This paper makes two related, but independent contributions.
First, we propose an alternative design for efficient scalable range locks that addresses the shortcomings of the existing
algorithm.
Our idea is to organize ranges in a linked list instead of a range tree.
Each node in the list represents an acquired range.
Therefore, conceptually, once a thread manages to insert its node into the list, it holds the range lock for that particular range.
While traversing a list to find the insertion point is less efficient than traversing a tree, 
the number of nodes in the list is expected to be relatively low, 
as it corresponds to the number of threads in the system accessing ranges.
At the same time, lists are known to be more amenable for non-blocking updates, 
since unlike a (balanced) tree, one needs to modify atomically just one pointer to update the list.
As a result, our list-based design does not require any lock in the common case.

Our second contribution is the discussion of applications for range locks in parallel software.
Our prime focus is on scaling the virtual memory management in the Linux kernel by
introducing a speculative mechanism into the \code{mprotect} operations.
As we observe, in certain cases handling \code{mprotect} calls results in modifying the metadata of the underlying VMA 
without changing the structure of \code{mm\_rb}.
For those cases, our mechanism acquires the range lock only for a relatively small (refined) range,
thus enabling parallel execution of \code{mprotect} operations on non-overlapping regions of virtual memory.
As it turns out, those are the common cases for applications that use the GLIBC memory allocator,
which is the default user-mode malloc-free allocator.
The latter employs per-thread memory arenas, which are initialized by mmaping a large chunk of memory and mprotecting 
the pages that are actually in use.
Those \code{mprotect} calls expand or shrink the size of the VMA corresponding to the set of pages with currently allocated objects,
which are exactly the cases that our speculative mechanism supports.

We note that the applicability of range locks extends beyond the virtual memory management subsystem.
As Kim et al.\ demonstrated recently~\cite{KKK19}, range locks can be used to optimize shared file I/O operations in a file system;
we believe that the range locks we present in this paper can be used as a drop-in replacement for
the implementation used in~\cite{KKK19}.
More generally, drawing from the original motivation behind the concept of range locks, 
the ideas presented in this paper appear to be a natural fit for parallel file systems;
we plan to experiment with such systems in the future work.
In addition, we argue that range locks can be highly useful in facilitating the design of scalable concurrent data structures.
As a concrete example, we discuss the design of a new skip list in which a range lock 
is used for scalable synchronization between threads applying concurrent operations on the skip list.
The new skip list is based on a well-known optimistic skip list by Helrihy et al.~\cite{HLL07}.
Instead of acquiring multiple locks during an update operation (potentially, as 
many as the number of levels in the skip list)~\cite{HLL07}, our design acquires one range only.
Beyond the potential performance benefits of reducing lock contention and the number of required 
atomic operations, our design eliminates the need for associating a (spin) lock with every node in the list, 
thus reducing the memory footprint of the skip list.

We have evaluated our ideas both in the user-space and kernel-space.
For the former, we implemented our list-based range locks and compared them 
to the tree-based range lock implementation that we ported from the Linux kernel into the user-space.
Our experiments confirm that the new range locks scale better and outperform existing range locks in virtually all evaluated settings.
Moreover, we show that the range lock-based skip lists perform significantly better when using our implementation of
range locks underneath, compared to the tree-based range lock implementation.
We also implemented the new range locks in the kernel, and evaluated them with Metis, a suite of map-reduce benchmarks~\cite{MMK10} 
used extensively for the scalability research of the Linux kernel~\cite{BWC10, CKZ12, KMK17, DK19b}.
When coupled with the speculative mechanism in \code{mprotect}, some Metis benchmarks run up to 9$\times$ faster
on the modified kernel compared to stock and up to 69$\times$ faster compared to the kernel that uses tree-based range locks.

\remove{
We present two variants of the range locks: one is for mutual exclusion, where threads acquire non-overlapping 
ranges for exclusive access, and another is adapted to reader-writer semantics,
where threads can acquire overlapping ranges as long as those threads require a read access.

While our work is heavily motivated by the Linux kernel use case, we stress that the proposed range locks are generic
and can be applied in other contexts, e.g., in parallel file systems, or as building blocks for concurrent data structures.
}

\input{related.tex}

\section{Existing Range Locks in the Kernel} 
\seclabel{existing-range-locks}
The existing implementation of range locks in the Linux kernel uses a range tree 
(based on red-black trees) protected by a spin lock~\cite{Kar13}.
To acquire a range, a thread first acquires the spin lock and then 
traverses the tree to find a count of all the ranges that overlap with (and thus, block) the given range.
For a reader-writer range lock, this count does \emph{not} include overlapping ranges belonging to other readers
(if the given acquisition is also for read)~\cite{Bue17}.
Next, the thread inserts a node describing its range into the tree, and releases the spin lock.
If at that point the count of blocking ranges is zero, the thread has the range lock and can start the critical section
that the lock protects.
Otherwise, it waits until the count drops to zero, which would happen when threads that have acquired blocking 
(i.e., overlapping) ranges exit their respective critical sections.
Specifically, when a thread is done with its range, it acquires the spin lock, removes its node from the tree and then traverses the tree, 
decrementing the count of blocking ranges for all relevant ranges, and finally releases the spin lock.

This range lock implementation has several shortcomings.
The most severe one is the use of a spin lock to protect the range tree.
This lock can easily become a bottleneck on its own even without the logical contention on ranges.
Note that every acquisition and release of the range lock
results in the acquisition and release of that spin lock.
Therefore, even non-overlapping ranges and/or ranges acquired for read have to synchronize 
using that same spin lock. 

Furthermore, while placing all ranges in the range tree preserves the FIFO order, it limits concurrency.
Assume that we have three exclusive acquisition requests for ranges coming in this order: A=[1..3], B=[2..7], C=[4..5].
While A holds the lock, B is blocked (it overlaps with A), and C is blocked behind B, but in practice, 
it could proceed as it does not overlap with A.
Finally, the existing range locks have no fast path, that is, even when there is a single thread acquiring a range,
it still would go through the same path of acquiring the spin lock, updating the range tree and so on.

The list-based range locks presented in this paper address all the aforementioned issues.
First, they only use a lock when fairness is concerned, i.e., to avoid starvation of threads trying
to acquire a range, but repeatedly failing to do so due to other threads that manage to acquire overlapping ranges.
In our experiments, this is an unlikely scenario, meaning that our range locks do not use any locks in the common case.
Second, list-based range locks can achieve a higher level of parallelism by allowing concurrent threads to acquire 
more (non-overlapping) ranges.
Considering the example above, for instance, while A is in the list, B waits until A finishes, 
but C can go ahead and insert its node into the list after A.
Finally, our design allows the introduction of a fast path, in which the 
range lock can be acquired in a small constant number of steps.
This path is particularly efficient for single-thread applications or multi-thread applications in which a range lock is acquired 
by one thread at a time.

We opted to use a linked list as an underlying data structure for the relative simplicity and amenability to concurrent updates of the former.
We note that, in general, a linear-time search provided by a linked list is less efficient than the logarithmic-time search provided
by a balanced search tree or a skip list.
In practice, however, this should not present an issue, as in all applications that we consider 
the number of stored elements (ranges) in the list is relatively small since it is proportional to the number of threads 
accessing concurrently the resource(s) protected by the range lock.
For the setting in which this assumption does not hold, we plan to investigate extending our design  to employ a skip list for more 
efficient search operations in the future.

\input{algo.tex}

\section{Refining Ranges in VM Operations} 
\seclabel{sec:refine-ranges}

\subsection{Background}
\seclabel{sec:vm:bkgrnd}
Operating systems provide processes with the virtual memory (VM) abstraction. It allows processes to assume they 
have access to all possible addressable memory, regardless of the actual underlying physical memory.
To keep track of how regions within a process's virtual memory map to actual physical memory pages (whether located in the main memory or swapped to disk), the Linux kernel uses the concept of Virtual Memory Area (VMA) structures~\cite{tor20}.
In practice, VMA is a data structure that defines a distinct contiguous 
region within the virtual memory address space using a start address and a variable length (multiple of a page size). 
The VMA metadata also includes other attributes, such as the 
mapping to physical memory, access permissions, pointers to neighboring VMA structures, etc. For each process, the 
Linux kernel stores all its associated VMA structures in a red-black tree (\code{mm\_rb}). A typical VM operation starts 
by querying \code{mm\_rb} with an address (provided as an input from the API caller) to find the enclosing VMA (if it exists). According to the nature of the operation, it may read or change some metadata of a VMA, split a VMA, merge 
two VMA structures, insert a VMA, delete a VMA, etc. 
Note that a single VM operation may perform several of these operations on one or more VMA structures, according to the given input address range. Moreover, splitting, merging, inserting and deleting VMA structures incur structural changes to the \code{mm\_rb}. To that end, operations that might modify VMA structures and/or\code{mm\_rb} (such as \code{mprotect}) acquire \code{mmap\_sem} for write, while operations that only read VMA’s metadata (such as the page fault handler) acquire \code{mmap\_sem} for read.

While the concept of range locks may appear, at a first glance, as a natural fit for 
synchronizing the access to regions of the shared virtual memory address space,
the task of applying those locks for this purpose in the Linux kernel is not straightforward due to mainly
two reasons: \emph{(i)}~the APIs of VM operations are oblivious to the underlying VMA structures, 
and rely on querying \code{mm\_rb} for this purpose; and \emph{(ii)}~a VM operation may end up performing structural changes to \code{mm\_rb} (and thus interfere with other concurrent VM operations accessing \code{mm\_rb}), and this is unknown a-priori. 

\sloppy
As a concrete example of the challenge of using range locks in VM operations, consider two calls:  \code{mprotect(0x100000, 65536, PROT\_NONE)} and 
\code{mprotect(0x180000, 65536, PROT\_READ)}.
If we naively protect only the range on which each call operates (i.e., [0x100000 .. 0x110000] and [0x180000 ... 0x190000]),
and those two ranges fall within the scope of the same VMA,
the two operations may simultaneously acquire range locks for the corresponding ranges, 
and  overwrite each other's updates to the metadata of that same VMA.
Moreover, if those calls result in a structural modification to \code{mm\_rb},
they would perform those modifications without synchronizing one with another.

To overcome these issues, one might always acquire the range lock for the full range whenever this lock 
is required in write mode.
This would, however, preclude any parallelism when a writer acquires the range lock, 
and in fact, is expected to perform worse than \code{mmap\_sem} (since the latter has a more efficient acquisition path). 


\subsection{\code{mprotect}}
\seclabel{mprotect}

\begin{figure}[t]
\centering
\subfloat[][Two adjacent VMA structures, with different protection flags.]{
\begin{adjustbox}{minipage=\linewidth,scale=0.5}
\includegraphics[width=1\linewidth, Clip=0 5cm 0cm 0.1cm]{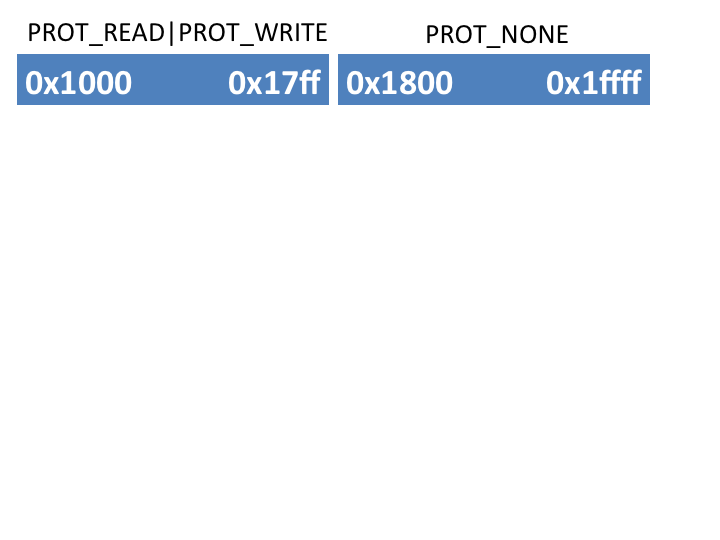}
\end{adjustbox}
}
\subfloat[][Same VMA structures after \code{mprotect(0x1800, 4096, PROT\_READ | PROT\_WRITE)} returns.]{
\begin{adjustbox}{minipage=\linewidth,scale=0.5}
\includegraphics[width=1\linewidth, Clip=0 5cm 0cm 0.1cm]{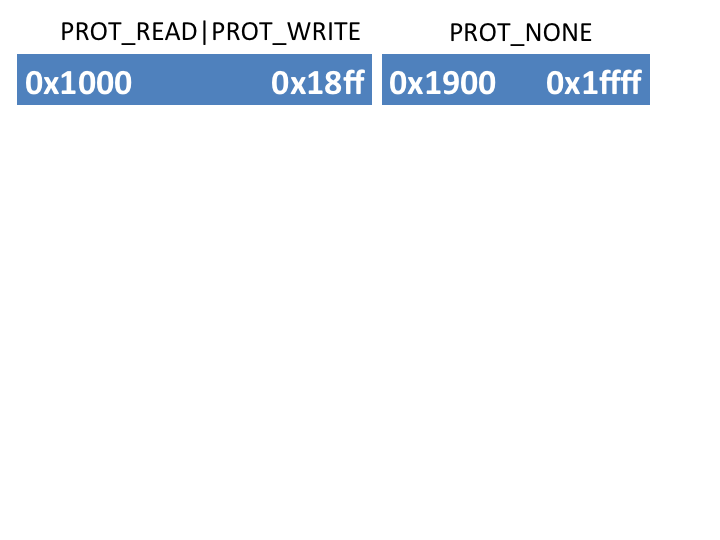}
\end{adjustbox}
}
\caption{Example for an \code{mprotect} operation changing VMA metadata without modifying the \code{mm\_rb} tree.}
\figlabel{fig:mprotect}
\end{figure}

By inspecting the implementation of various VM operations~\cite{tor20},
we notice that they do not always end up modifying \code{mm\_rb}.
For instance, consider the case when there are two neighboring VMA structures describing two contiguous memory regions 
with different protection flags, 
and \code{mprotect} is called on the area at the head of the second VMA (or the tail of the first VMA), with 
protection flags identical to the flags of the other VMA (see \figref{fig:mprotect}).
In that scenario, the boundaries (i.e., the metadata) of the involved VMA structures are changed, but the structure of 
\code{mm\_rb} remains unchanged.
As mentioned in the Introduction, this case is common in the GLIBC memory allocator.
Consequently, for the cases where \code{mm\_rb} does not change, we devise a speculative approach,
in which the range lock is optimistically acquired only for the relevant part of the VM address space. We note that when a VM operation needs to modify \code{mm\_rb} (e.g., when \code{mprotect} splits a VMA into two, thus it needs to create 
a node corresponding to the new VMA and insert it into \code{mm\_rb}), acquiring the range lock for the entire range 
is the only available option to synchronize correctly with other operations traversing the \code{mm\_rb} tree.

Listing~\ref{alg:mprotect} provides the pseudo-code for the \code{mprotect} operation with the integrated
speculative mechanism. The intuition behind our speculative approach is that if we are able to decide whether 
the \code{mprotect} operation will end up modifying \code{mm\_rb} before the \code{mprotect} applies its changes, 
then it is safe to lock only the respective range; otherwise, if we discover that \code{mm\_rb} needs to be modified, 
we restart the \code{mprotect} operation after acquiring the full range (for write).
The latter action prevents other concurrent speculative operations from running and potentially reading inconsistent \code{mm\_rb} 
while it is being modified.
To this end, we augment the major memory management structure in the Linux kernel (\code{mm}) with a sequence number.
This number is incremented every time a range lock acquired for the full range in write mode is released.
We use the sequence number to detect whether the \code{mm\_rb} has changed during the 
speculative operation as described below.

The first step of the \code{mprotect} operation is to locate the relevant VMA given the input address and size.
Therefore, we first acquire the range lock in read mode for the input range.
This ensures that the structure of the underlying \code{mm\_rb} would not change while \code{find\_vma()} is
running, since we make sure that \code{mm\_rb} only changes under the range lock acquired in write mode for the entire range.
(As its name suggests, \code{find\_vma()} traverses \code{mm\_rb} searching for the VMA that contains the given address,
or more precisely, searching for the first VMA whose end address is larger than the given address).
Note that since the range lock is acquired in read mode, this step may run in parallel with other
speculating operations (or any other operation that acquires a range lock in read mode).
After locating the VMA, we unlock the range lock, and lock it again, this time in write mode and 
with the range adjusted to span the entire VMA
(plus some small extra space, as we explain below).
Note that during the time the range lock is not held, \code{mm\_rb} may change and, in particular, the VMA
returned by \code{find\_vma()} might not be valid anymore. 
We use a sequence number mentioned above to detect this scenario.
Specifically, we read the sequence number right before dropping the read range lock and compare it to the
number read right after acquiring the write range lock.
If those numbers differ (or the boundaries of the found VMA have changed), 
the speculation fails, and we restart the \code{mprotect} operation from the beginning.
We note that it is trivial to limit the number of retries, although we do not do that in our prototype implementation.

In case the speculation can proceed, we continue with the operation by going through the logic of 
identifying the required changes to the VMA(s) involved in the given \code{mprotect} operation.
If this logic identifies that the changes require a structural modification to \code{mm\_rb}, the speculation fails, 
the write range lock is dropped, and the \code{mprotect} operation is restarted by acquiring the write range lock for the full range.
Otherwise, the \code{mprotect} operation completes while holding the write range lock for the relevant range only, thus
allowing parallelism with other \code{mprotect} operations and/or operations that acquire the range lock for read (e.g., page faults 
discussed in the next section).

We are left to describe one subtle detail of determining the size of the range for the write acquisition during speculation.
We note that it is not enough to lock only the underlying VMA of the given \code{mprotect} operation.
This is because as discussed in Section~\secref{sec:vm:bkgrnd}, 
two \code{mprotect} operations on neighboring VMA structures can change the metadata of one another concurrently,
thus creating a race condition.
To avoid this situation, we set the range of the write range lock acquisition to the underlying VMA plus a page (4096 bytes)
from each side of the VMA.

\StartLineAt{1}
\begin{figure}[t!]
\begin{lstlisting}[language=Python, caption=Simplified pseudo-code for the speculative \code{mprotect} implementation., label=alg:mprotect, escapechar=|,deletendkeywords={next}, commentstyle=\color{blue}]
mprotect(__u64 addr, size_t size, int prot_flags):
	__u64 start = addr
	__u64 end = addr + size
	bool speculate = true	

	while true:
		if speculate: range_read_lock(range_lock, start, end)
		else:		     range_full_write_lock(range_lock)
	
		vm_area_struct *vma = find_vma(addr)
	
		if speculate:
			__u64 seq_number = mm->seqnumber
			__u64 aligned_start = vma->start - 4096
			__u64 aligned_end = vma->end + 4096
		
			range_read_unlock(range_lock)	
			range_write_lock(range_lock, aligned_start, aligned_end)
		
			if seq_number !=  mm->seqnumber or 
				aligned_start != (vma->start-4096) or 
				aligned_end != (vma->end+4096):
				##validation failed, retry
				range_write_unlock(range_lock);
				continue
		
		##apply mprotect logic
		...	
		if speculate and will perform structural modification:
			range_write_unlock(range_lock)
			speculate = false
			continue
		...
		release_write_unlock(range_lock)
		return
\end{lstlisting}
\end{figure}

While the speculative mechanism described in this section is presented in the context of \code{mprotect},
we note that a similar mechanism can be employed in other operations as well.
For instance, \code{mmap}, \code{munmap} and \code{brk} all start from calling \code{find\_vma} (or a similar function), 
during which the range lock can be held in the read mode.
Those operations, however, typically (but not always) end up modifying \code{mm\_rb}, and thus would 
need to drop the read range lock and acquire the write range lock for the entire range.
Thus, the speculative approach would shorten the time during which the write range lock is held at the cost
of an extra (read) range lock acquisition.
Evaluating the effect of this speculation is left for future work.

\subsection{Page Faults}
Page fault interrupts access the VM subsystem to identify whether the address that triggered the fault is allowed to be accessed.
They do so by locating the appropriate VMA (by calling the same \code{find\_vma()} function) and then handling the fault based on that VMA's
metadata (such as protection flags).
Since the page fault routine only queries the metadata of VMA structures (but does not change them), it acquires the range lock in read mode.
The original patch that introduced range locks into the Linux kernel, however, does all the acquisitions, including the one in the 
page fault routine, for the full range~\cite{Bue18}.

We observe that the page fault routine accesses only the metadata of the VMA returned by \code{find\_vma()}.
Therefore, it is straightforward to refine the range of the lock acquisition to contain only the given address (in our implementation, we
lock the range of a page size). 
We note that any modification to \code{mm\_rb} is done while holding the write range lock for the full range, while 
any modification to VMA metadata is done while holding the write range lock (at least, according to \secref{mprotect}) 
that covers the range being modified.
Therefore, the refinement of the range of the lock acquired in page faults is safe.
Furthermore, note that this refinement alone is not expected to improve the scalability of the VM subsystem,
because the range lock is acquired in read mode, similarly to the original \code{mm\_sem}.
However, when coupled with the speculation in \code{mprotect}, page fault interrupts can now lock and access VMA structures 
in parallel with some (or at least part) of the \code{mprotect} operations.

\section{Range Lock-based Skip Lists}
\label{sec:skiplist}
In this section, we show how range locks can be used to coordinate concurrent accesses to a skip list.
We base our design on the optimistic skip list by Herlihy et al.~\cite{HLL07}.
In the original design, each node is associated with a spin lock.
Search operations are wait-free, and in particular do not acquire any locks.
Update operations start by searching the list for the given key, locking all relevant nodes (we elaborate on that below) and
validating that the list has not changed in a way that precludes completing the operation (e.g., the node we want to 
delete is still in the list), perform the required update (removing the node from the list, or inserting a new node), and finally
unlock all the acquired locks. 
If the validation above fails, the operation releases all the locks it has acquired, and restarts.

When replacing the per-node spin lock with a single range lock, we maintain the same properties.
In particular, the search operations are still wait-free, which is important for read-dominated workloads.
The major change is in the locking protocol.
The original optimistic skip list acquires node-level locks for all the predecessors of the node returned by search (in case of a remove operation)
or of the node with a key larger than the given key (in case of an insert operation).
Note that each node has between 1 and N predecessors, where N is the number of levels in the skip list, 
and thus the locking protocol consists of between 1 and N lock acquisitions.
In addition, remove operations acquire the lock of the target node to be deleted, adding one more lock
acquisition to the locking protocol.
With range locks, we always need to acquire one range only.
For inserts, the range is the interval between the key of the predecessor at the highest level (at which 
the new node will be inserted) and the target key (to be inserted). 
For removes, the range is defined from the key of the predecessor at the highest level to the target key (to be removed) plus 1;
the latter is to avoid races with inserts that may attempt to update pointers in the to-be-deleted node.

We note that beyond the conceptual simplicity and the potential performance benefits stemming from the fact that each 
operation acquires at most one (range) lock, the range lock-based skip list has a smaller memory footprint than its original 
lazy counterpart. 
This is due to elimination of spin locks associated with every node in the skip lists.
As the number of nodes in skip lists is typically (much) larger than the number of concurrent threads updating the skip list,
this may translate into significant memory savings.

\section{Performance Evaluation}
\label{sec:eval}

\subsection{User-space}
In this section, we evaluate our linked list-based range locks using two user-space applications.

We start with ArrBench, a microbenchmark that we developed in which threads access a range of slots of a shared array for either read or write. This benchmark 
allows us to assess the performance of our range locks in different contention scenarios.
Array slots are padded to the size of a cache line.
In read mode, a thread reads the values stored in each slot in the given range, while for write a thread increments the 
value stored in each slot by $1$.
Each operation acquires a range lock for the corresponding range, and in the corresponding access mode (read or write).
Between operations on the array, each thread performs some (non-critical) work, emulated by a variable number of no-op operations.
The number of no-op operations is chosen uniformly randomly from the given range (2048 in our case). 
We set the size of the array (i.e., the number of slots) to $256$.



To simulate various levels of contention and possible usage scenarios for range locks, 
we created three variants of the ArrBench: in the first variant, each thread acquires the entire range of the array.
In the second variant, each thread acquires a non-overlapping range calculated by dividing the size of the array by the number of threads.
Note that in this variant, threads do not conflict on the ranges they acquire.
Furthermore, in order to keep the amount of work (i.e., the number of slot accesses) performed under the range lock 
the same independent of the number of threads, in this variant only, threads traverse the corresponding portion 
of the array the number of times equal to the number of threads.
In other words, when this variant is run with one thread, that thread would traverse the entire array once for every 
acquisition of the range lock; when run with two threads, each of the threads would traverse half of the array twice for every 
acquisition of the range lock, and so on.
Finally, in the third variant, each thread picks random starting and ending points from the range defined by the size of the array\footnote{We select 
starting and ending points randomly modulo the size of the array, and switch if the former is larger than the latter.},
acquires the range lock with that range, and performs one traversal of corresponding slots.

We implemented the mutex and \rw variants of the range lock described in the paper
(without the fast path and fairness optimizations -- we leave the evaluation of those for future work). 
We denote those variants as \code{list-ex} and \code{list-rw}, respectively.
We ported two implementations of range locks found in the kernel into the user-space,
one found in the Lustre file system (denoted as \code{lustre-ex}) and another recently 
proposed by Bueso~\cite{Bue17} (denoted as \code{kernel-rw}).
As mentioned earlier, the latter is a \rw version of the former.
In the user-space experiments, we used a simple test-test-and-set lock to implement a spin lock protecting the range tree 
in \code{lustre-ex} and \code{kernel-rw}.
We note that the Linux kernel uses a slightly more sophisticated spin lock implementation~\cite{linux-locks, DK19},
however, this detail is insignificant in our context\footnote{To confirm that, we tried a different lock and observed similar relative performance results.}.
In addition, we implemented the recent proposal for range locks by Kim et.al.~\cite{KKK19}.
Those locks were proposed in the context of pNOVA, a variant of a non-volatile memory file system, hence
we denote this version of range locks as \code{pnova-rw}.
As described in Section~\ref{sec:related}, \code{pnova-rw} operates with a present number of segments, each of a preset size~\cite{KKK19};
in our experiments we set this number to $256$ segment, spanning one array slot each.
We also experimented with other number of segments, spanning multiple slots; although the results were quantitatively different,
they lead to similar conclusions.

We ran the experiments on a system with two Intel Xeon
E5-2630 v4 sockets featuring 10 hyperthreaded cores each
(40 logical CPUs in total) and running Fedora 29. We
did not pin threads to cores, relying on the OS to make its
choices. 
We also disabled the turbo mode to avoid the effects 
of that mode (which may vary with the number of threads) 
on the results. 
We vary the number of threads between $1$ and $40$,
as well as the mix of operations performed by each thread ($100\%$ reads,
$80\%$ reads and $20\%$ writes, and $60\%$ reads and $40\%$ writes).
The results for the $80\%$ reads workload were similar to the $60\%$ reads workload and thus omitted.
Each reported experiment has been run 5 times in
exactly the same configuration. 
Presented results are the
mean of throughput results reported by each of those 5 runs,
where throughput is calculated based on 
the total number of operations performed by all the threads running for ten seconds. 
The standard deviation of nearly all results is less than $3\%$ of the mean.

\begin{figure}[!tb]
\subfloat[][100\% reads]{\includegraphics[width=0.5\linewidth]{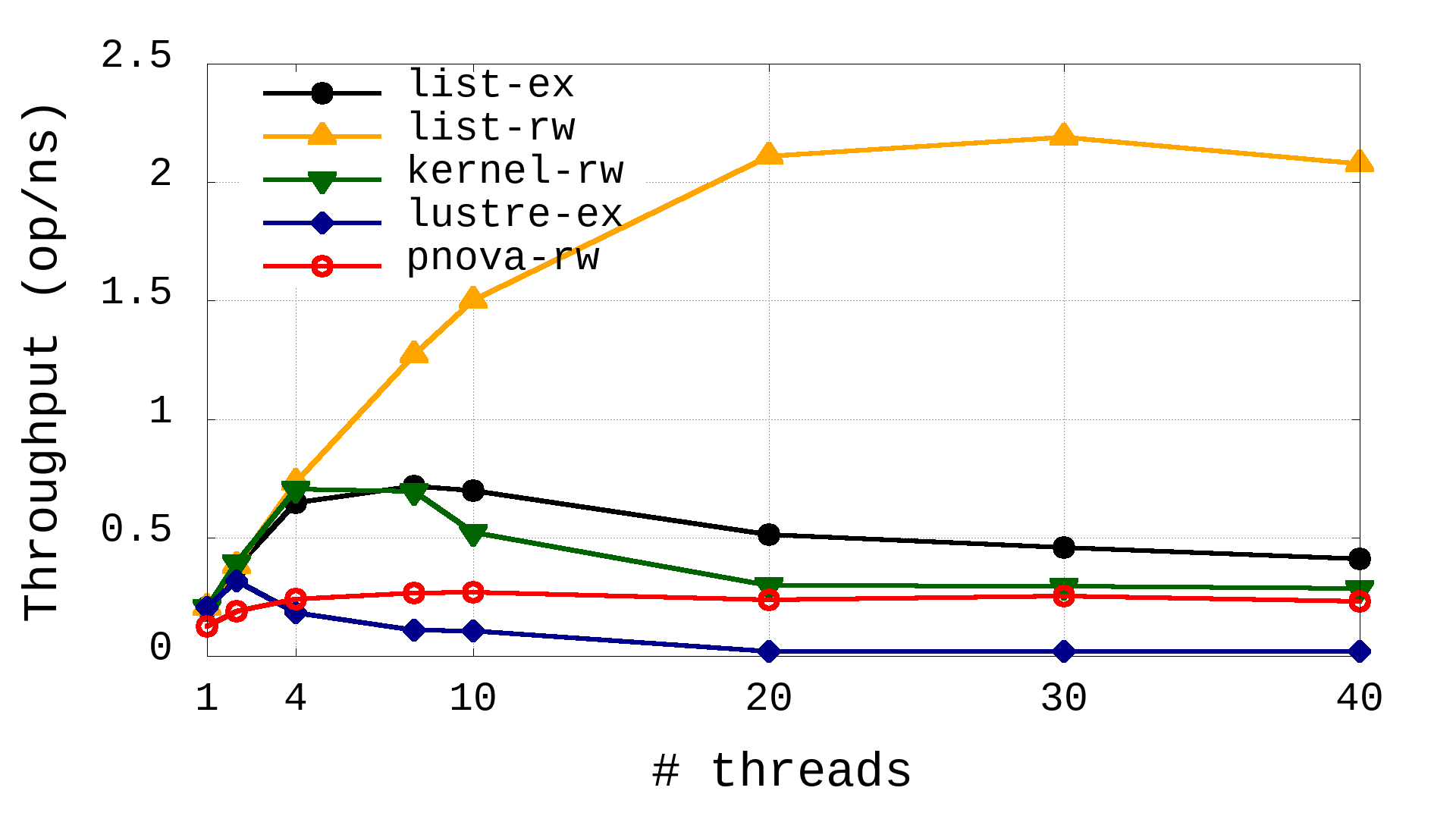}}
\subfloat[][60\% reads]{\includegraphics[width=0.5\linewidth]{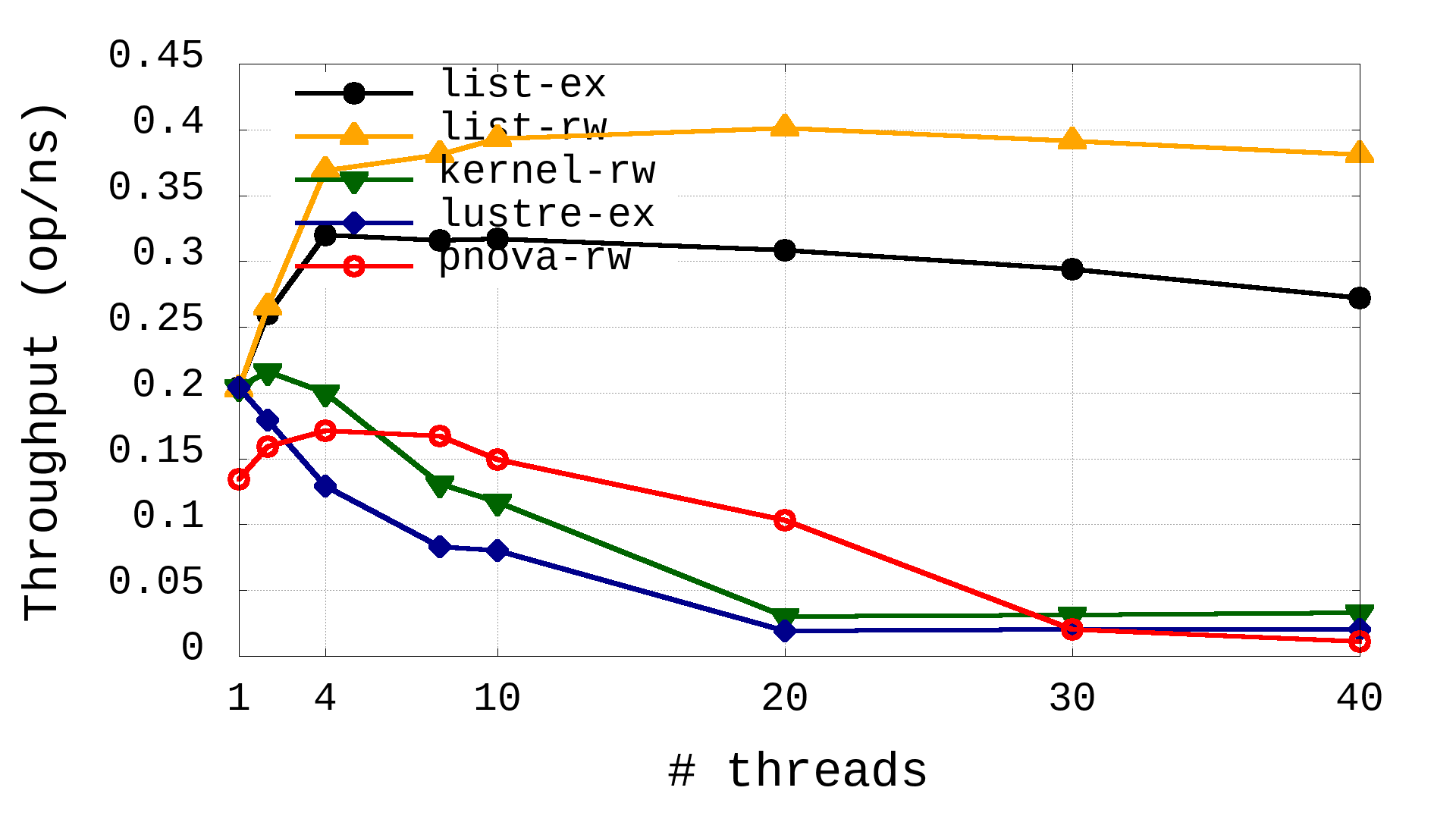}}
\vspace{-1.2em}
\subfloat[][100\% reads]{\includegraphics[width=0.5\linewidth]{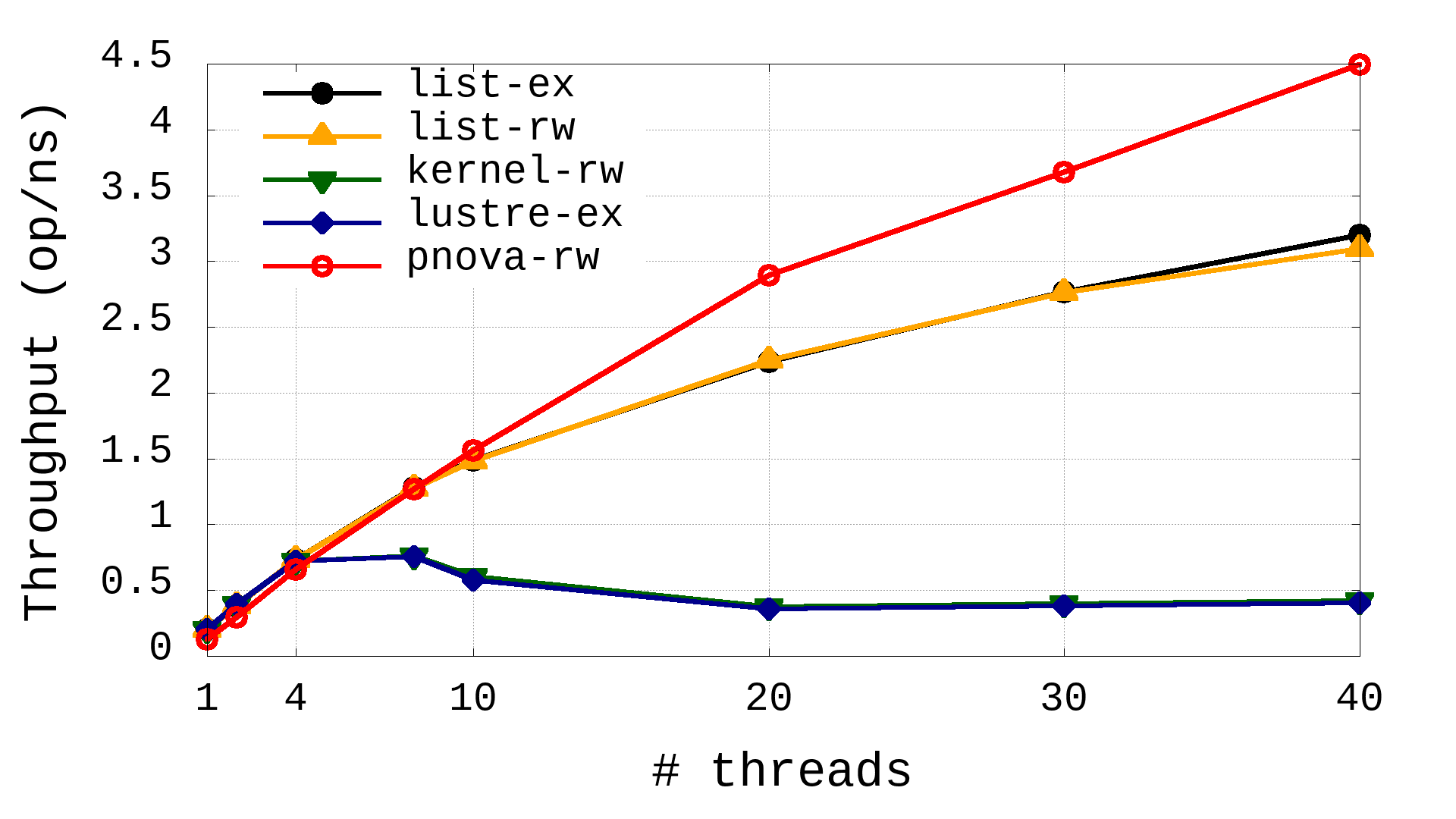}}
\subfloat[][60\% reads]{\includegraphics[width=0.5\linewidth]{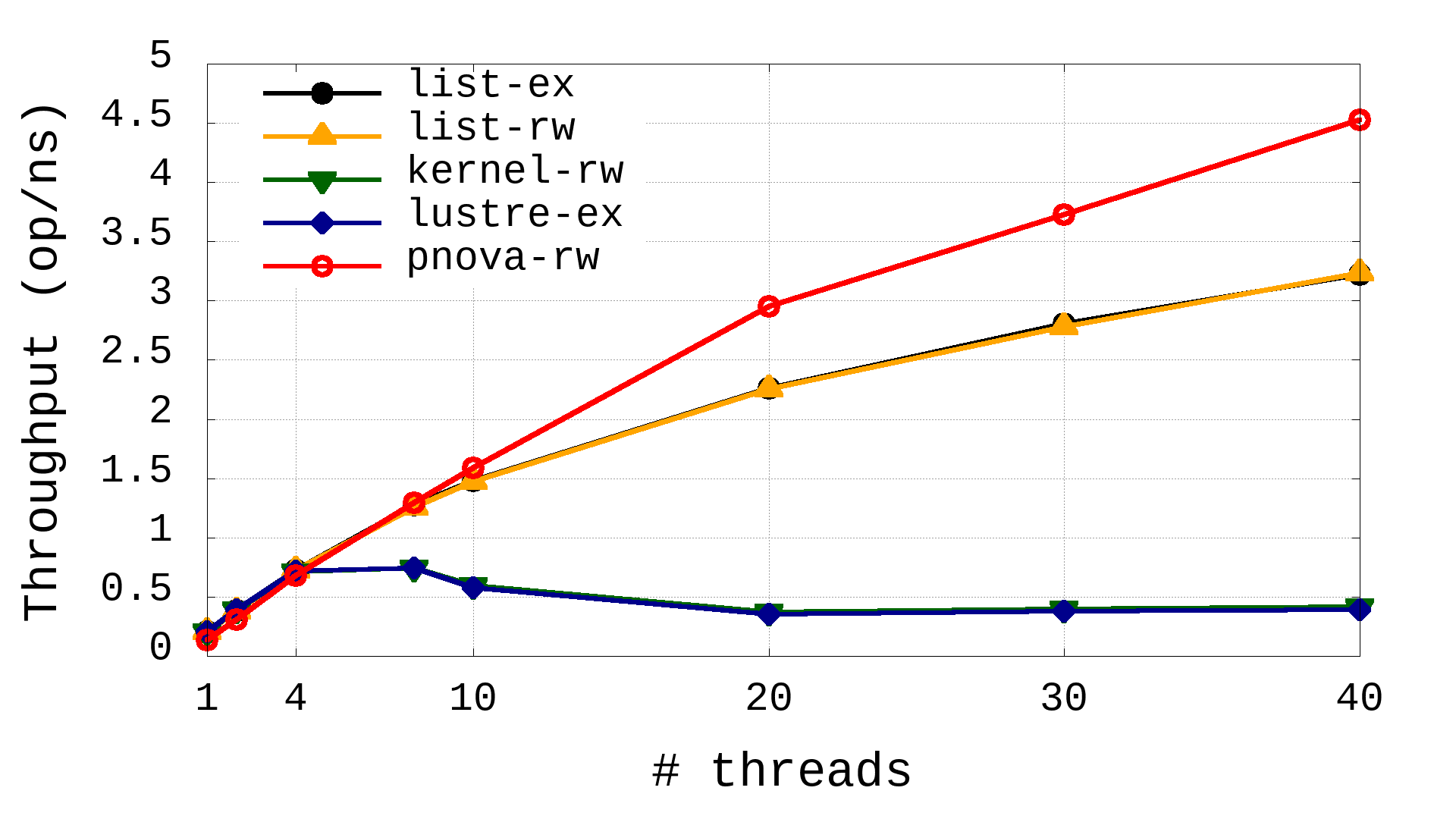}}
\vspace{-1.2em}
\subfloat[][100\% reads]{\includegraphics[width=0.5\linewidth]{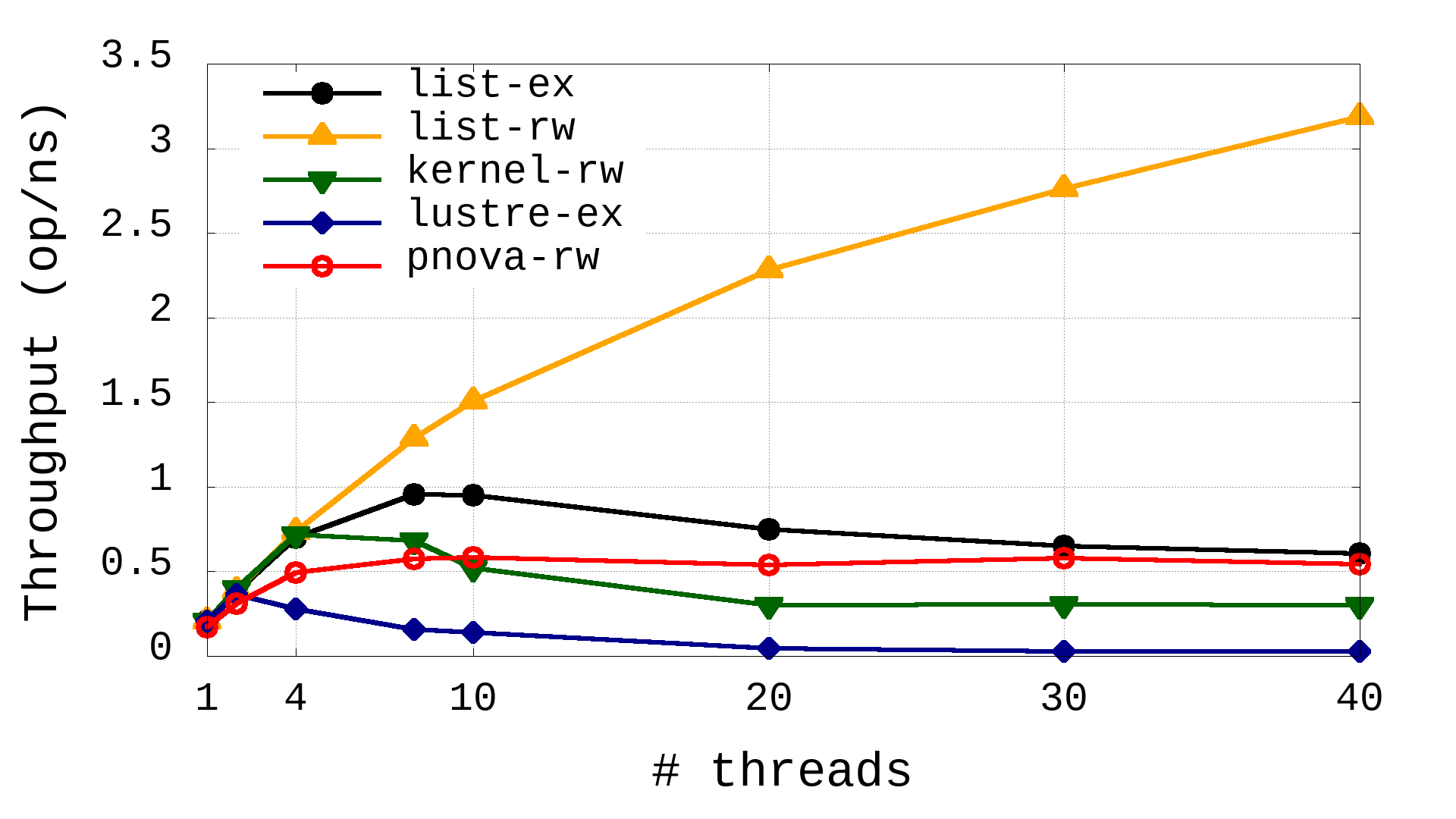}}
\subfloat[][60\% reads]{\includegraphics[width=0.5\linewidth]{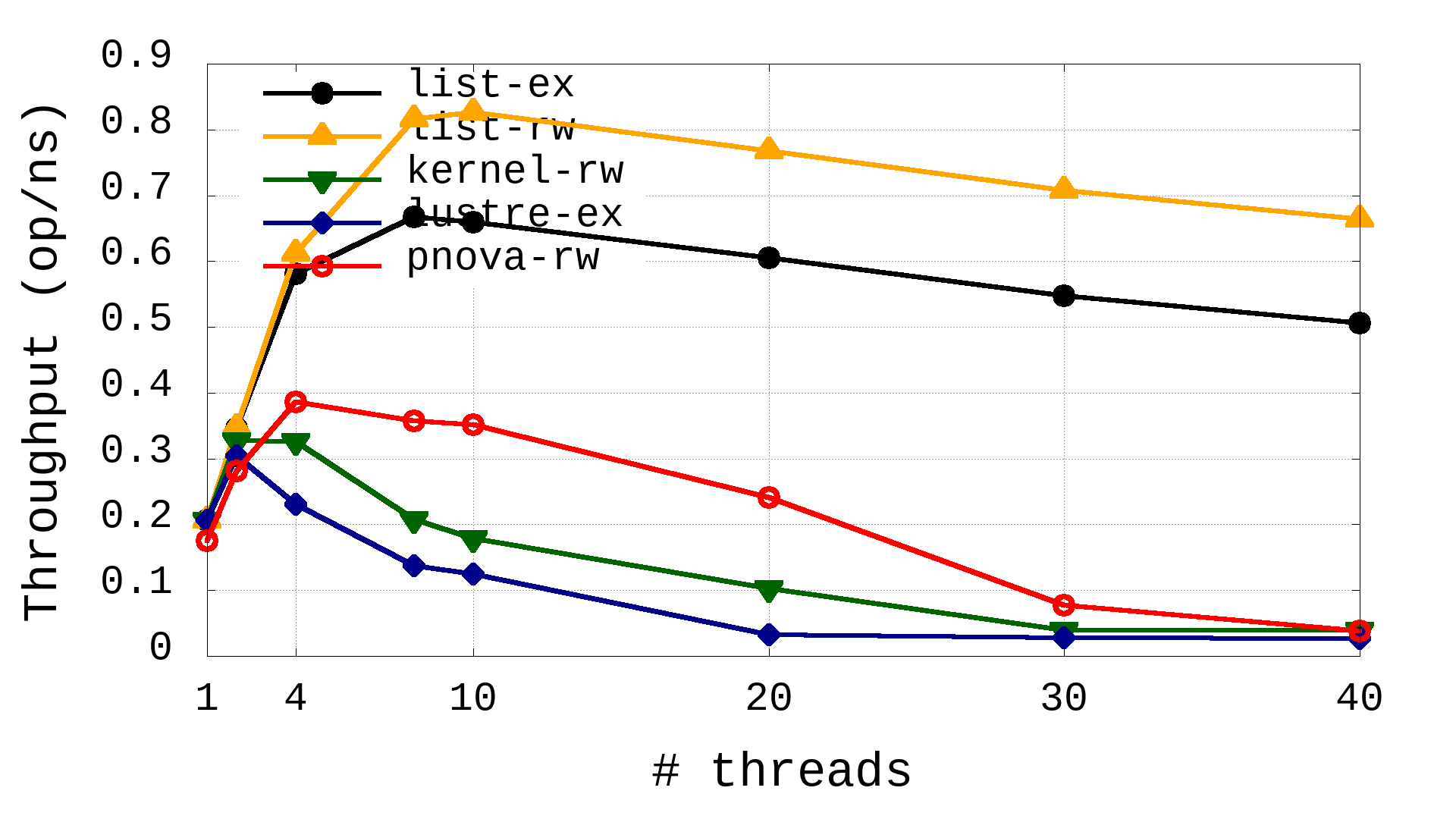}}
\caption{Throughput for the ArrBench microbenchmark, where all threads acquire the entire range (first row), 
threads acquire non-overlapping ranges (second row) and threads acquiring random ranges (third row).}
\figlabel{fig:arrbench-1}
\end{figure}

The results for the first variant of ArrBench, in which threads acquire and access the entire range, are shown in \figref{fig:arrbench-1}~(a) and~(b).
The \code{lustre-ex} variant does not scale at all, it allows only one thread to traverse the array at a time as it does not support \rw\ semantics. 
Moreover, all threads contend heavily on the spin lock protecting the range tree structure.
This is not the case for the \code{list-ex}, where the fact the threads perform non-critical work
without a range lock helps it to scale for low thread counts.
In fact, in most cases \code{list-ex} performs better than \code{kernel-rw}, even though the latter allows readers run concurrently.
Once again, the spin lock protecting the underlying range tree plays detrimental role in the performance of \code{kernel-rw}.
The \code{pnova-rw} variant also does not scale due to the high lock acquisition latency 
(acquiring this lock for the entire range requires acquiring all the underlying segment reader-writer locks).
At the same time, \code{list-rw} does not use locks in the common case, and shows scalability across most thread counts.

The results for the second variant of ArrBench, in which each thread acquires a non-overlapping part of the range, are shown in \figref{fig:arrbench-1}~(c) and~(d).
Note that the maximum number of concurrent range accesses is equivalent to the number of threads depicted on the 
x-axis, which determines the size of the list (or the tree) in the corresponding range lock implementation.
In theory, in this case the total throughput should scale with the number of threads for every range lock, as threads never compete 
for the same range (regardless of the access mode).
In practice, however, all range locks scale almost linearly up to a small number of threads ($4$--$8$).
Beyond that, the contention on the spin lock in \code{lustre-ex} and \code{kernel-rw} degrades the performance of those variants.
\code{list-ex} and \code{list-rw} lack a single point of contention, and manage to scale, albeit less than linearly, across all thread counts.
\remove{
Notably, \code{list-rw} performs slightly better than \code{list-ex}, and both achieve higher throughput when the workload contains more updates.
We believe that both phenomena stem from the same cause -- in \code{list-rw}, writers (and readers) perform a validation step that introduces
extra work, which serves as a back-off mechanism that helps to reduce contention on list node pointers.
This effect of extra work diffusing contention and improving overall performance has been noted before~\cite{CDHHKMM13}.
}
\code{pnova-ex} tops the charts as in this workload none of its underlying segment reader-writer locks is contended.

When considering the results for the third variant of ArrBench, in which each thread acquires a random part of the range (see \figref{fig:arrbench-1}~(e) and~(f)), 
one can note a mix of behaviors seen in the previous two variants.
Overall, \code{lustre-ex} does not scale, \code{kernel-rw} scales up to a small number of threads, 
while \code{list-ex} either slightly better than (in read-only workload) 
or significantly outperforms (when workloads include writes) \code{kernel-rw}, despite providing only exclusive access to each range.
\code{pnova-ex} performs poorly as its underlying reader-writer locks are once again contended.
At the same time, \code{list-rw} provides superior performance across all workloads, scaling better than any other variant.

Next, we used the Synchrobench benchmark~\cite{Gra15} to evaluate the performance of new skip lists that employ
range locks to synchronize concurrent access, as discussed in Section~\ref{sec:skiplist}.
We compare three variants: the original optimistic skip list~\cite{HLL07} (provided in Synchrobench, denoted as \code{orig}), 
and two variants of our new skip list that uses a range lock, one built on top of the Lustre range locks 
(denoted as \code{range-lustre}) and another on top of the exclusive list-based range lock presented in Section~\ref{sec:exclusive}
(denoted as \code{range-list}).
As it is not clear how one should set the number and the size of segments in pNOVA range locks, 
we do not include that lock in the evaluation of skip lists.

\figref{fig:skiplist-80} shows the results for the typical set workload composed of 80\% find and 20\% update operations 
(split evenly between inserts and removes); the key range is 8M, and 4M keys are randomly selected and 
inserted into the skip list before each experiment.
We report the mean throughput after repeating each experiment 5 times 
(here as well the standard deviation is less than $3\%$ of the mean for nearly all data points).
The results show that \code{range-list} performs similarly to \code{orig},
even though the former is simpler and consumes less memory as it does not use a lock per skip list node.
\code{range-lustre} tracks both versions at lower thread counts. 
Once thread counts grow, however, the contention on its internal spin lock increases, and as expected, its performance drops to less 
than half of the other two variants. This workload demonstrates that the increased concurrency allowed by \code{range-list} outweighs the linear complexity of the linked list, 
in contrast with the logarithmic complexity of \code{range-lustre}'s range tree. 

\begin{figure}[t]
\includegraphics[width=0.95\linewidth]{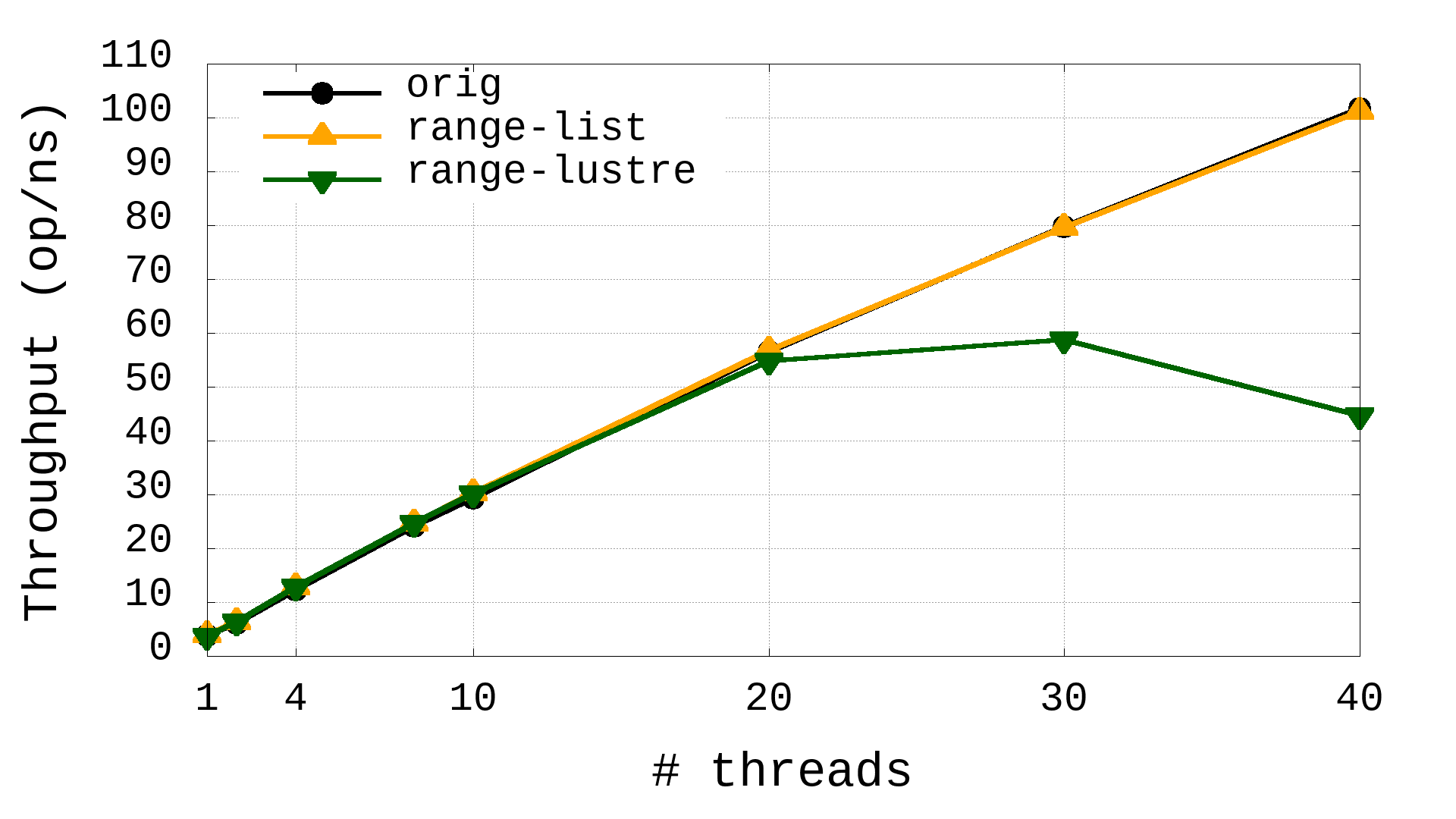}
\caption{Throughput for the skip list benchmark.}
\figlabel{fig:skiplist-80}
\end{figure}

\subsection{Kernel-space}
For the kernel-level experiments, we compared the stock version (4.16.0-rc2) with the one that has 
\code{mm\_sem} replaced with a range lock.
For the latter, we used the patch by Bueso~\cite{Bue18}; we call this variant \code{tree-full} as
it always acquires the range lock for the full range.
Based on this patch, we replaced the range lock implementation with the reader-writer linked list-based one
described in this paper; we call this variant \code{list-full}.
Furthermore, we refined the ranges of the acquired range locks as described in \secref{sec:refine-ranges}.
We refer to the variants with refined ranges as \code{tree-refined} and \code{list-refined}, respective of the range lock 
implementation used by each.
All the variants were compiled in the default configuration.

\begin{figure*}[!th]
\subfloat[][wr]{\includegraphics[width=0.33\linewidth]{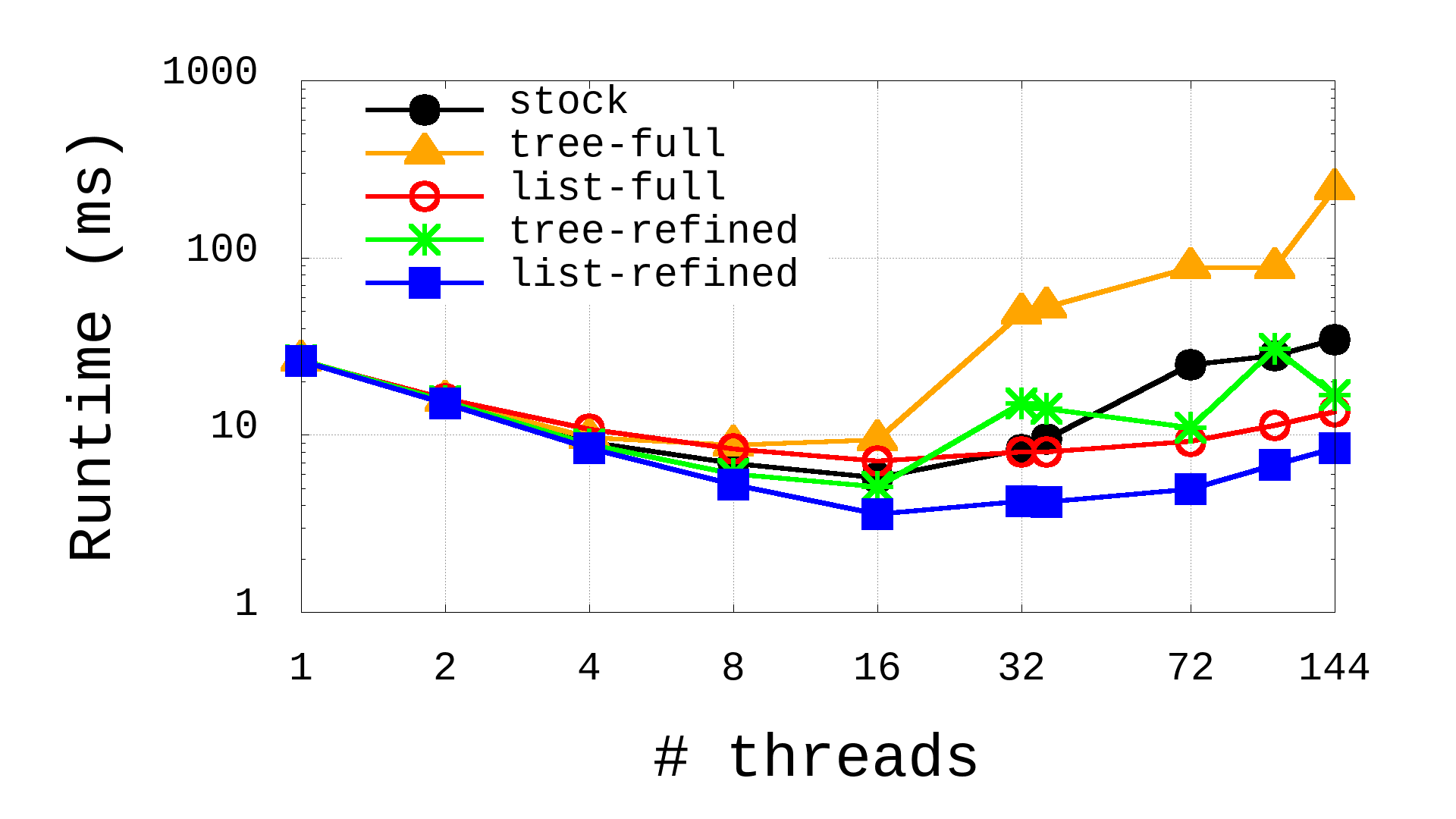}}
\subfloat[][wc]{\includegraphics[width=0.33\linewidth]{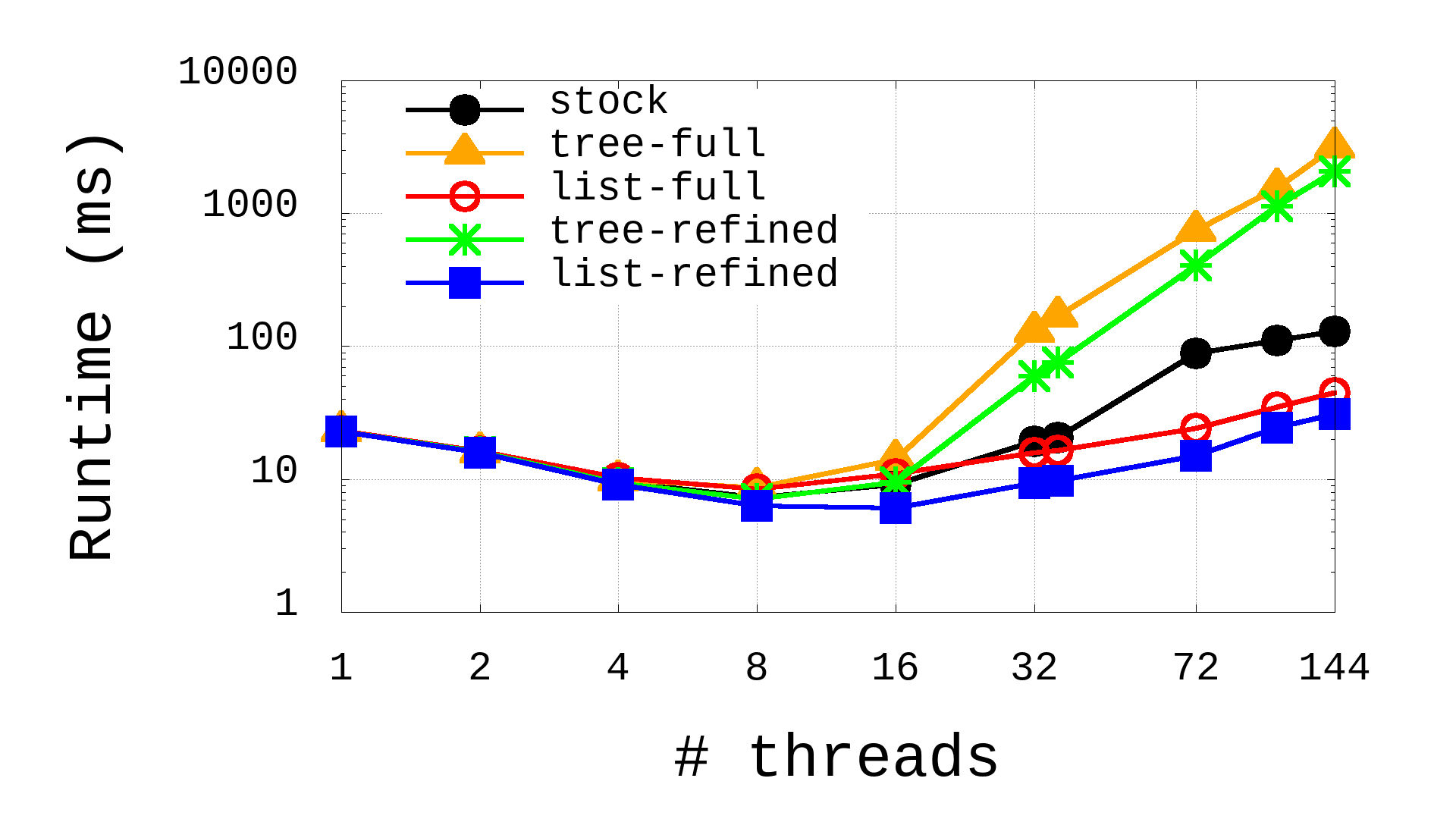}}
\subfloat[][wrmem]{\includegraphics[width=0.33\linewidth]{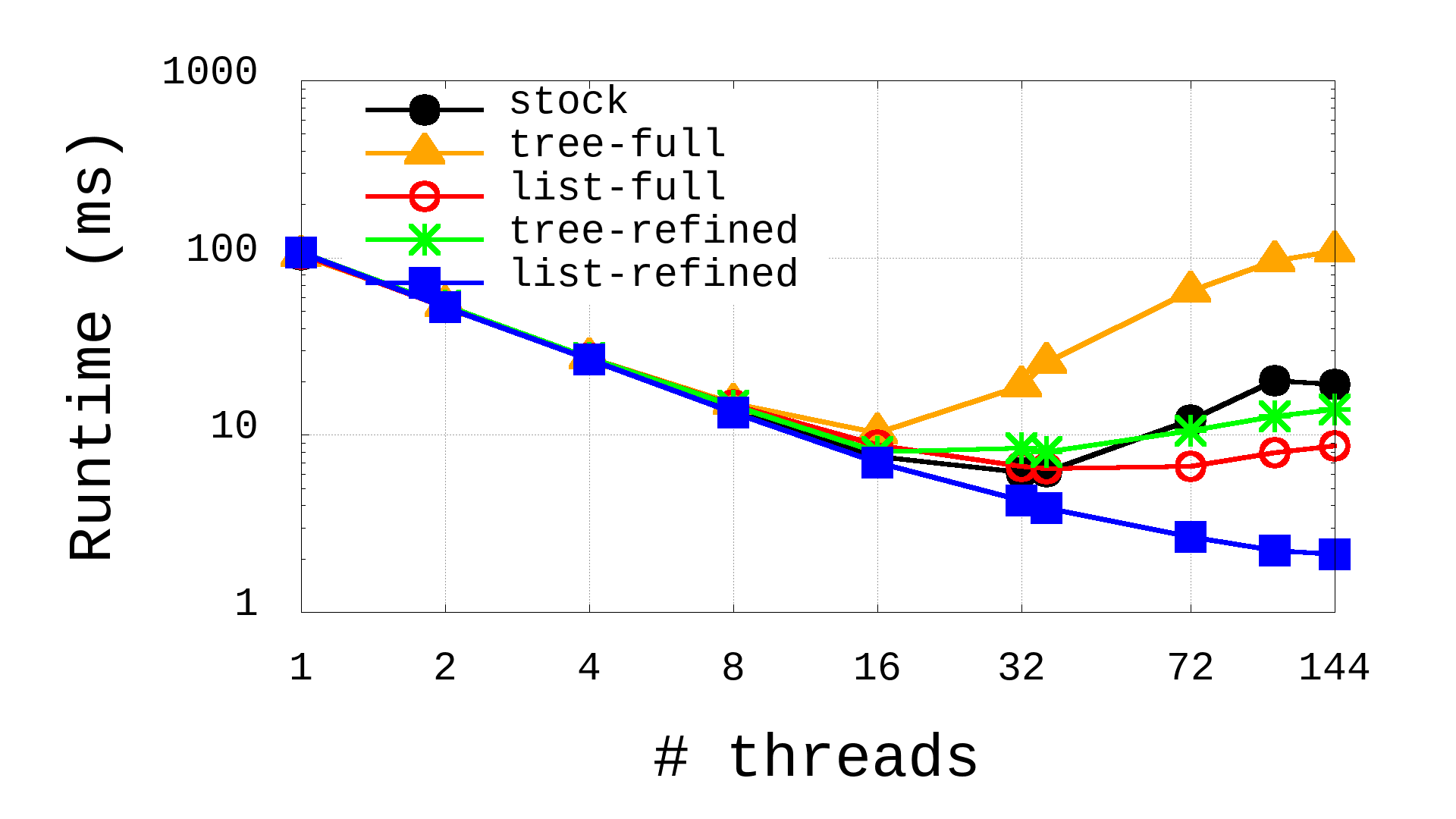}}
\caption{Runtime for Metis benchmarks.}
\figlabel{fig:metis-tput}
\vspace{-4mm}
\end{figure*}

\begin{figure*}[!th]
\subfloat[][wr]{\includegraphics[width=0.33\linewidth]{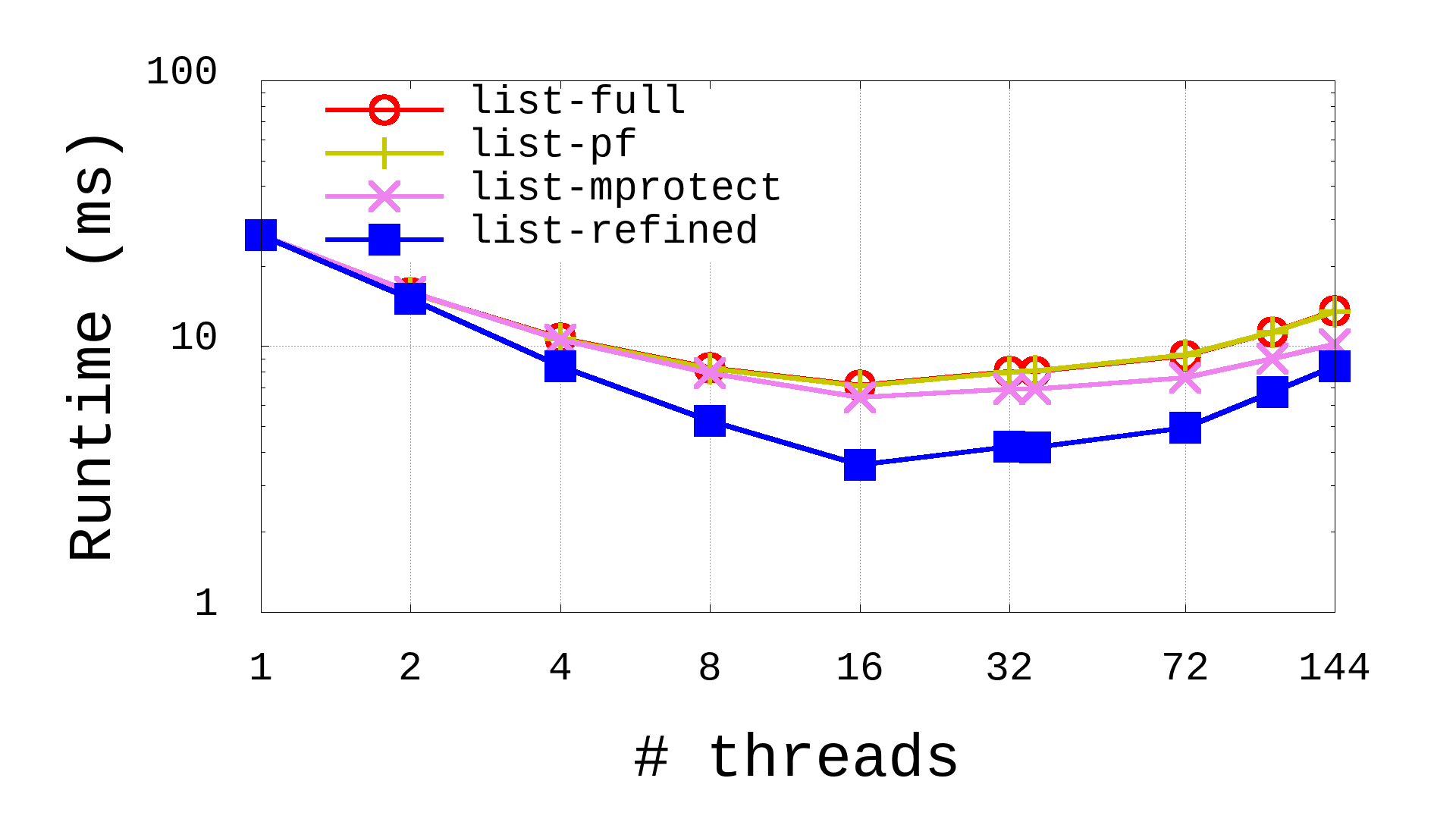}}
\subfloat[][wc]{\includegraphics[width=0.33\linewidth]{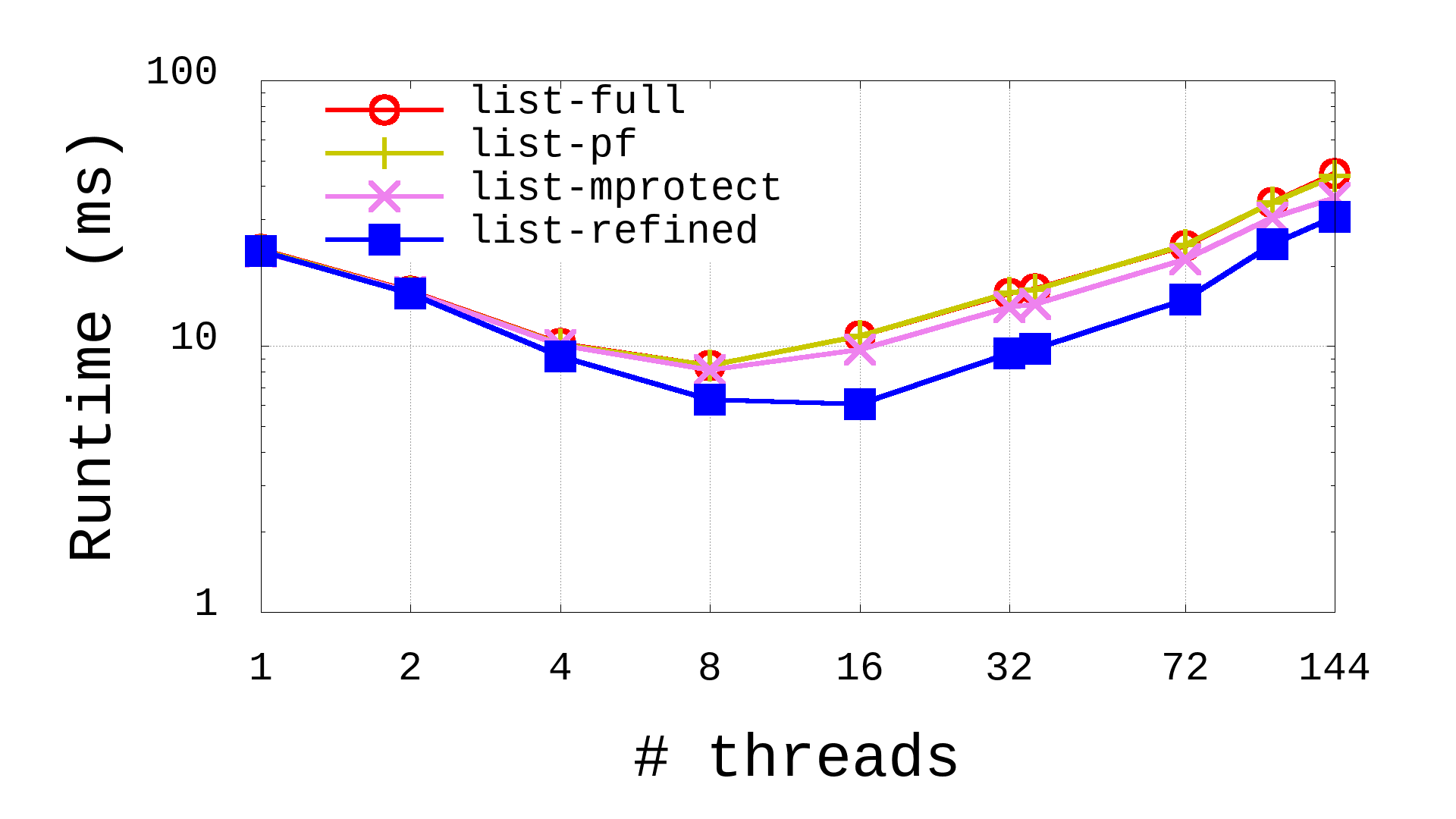}}
\subfloat[][wrmem]{\includegraphics[width=0.33\linewidth]{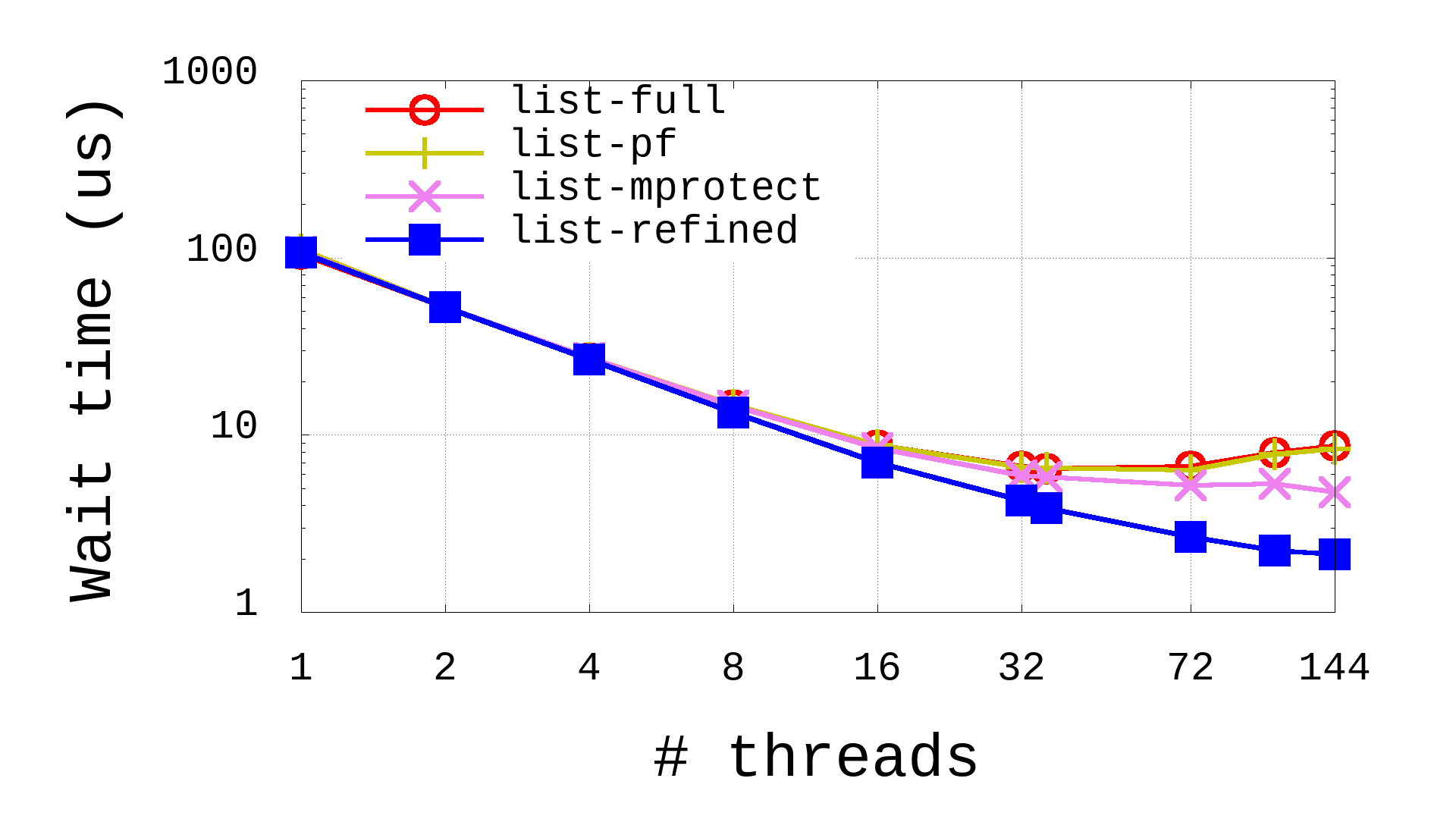}}
\caption{Breakdown of the impact of refining the range in list-based range lock variants.}
\figlabel{fig:metis-tput-refinement}
\end{figure*}

We ran the experiments on a system with four Intel Xeon
E7-8895 v3 sockets featuring 18 hyperthreaded cores each
(144 logical CPUs in total). 
Like for user-space experiments, we do not pin threads to cores and disable the turbo mode.
For our evaluation, we used Metis, an open source MapReduce library~\cite{MMK10}, known for 
stress-testing the VM subsystem through the mix of VM-related operations (such as page-faults, 
\code{mmap} and \code{mprotect})~\cite{KMK17}.
Each experiment was repeated $5$ times, and we report the mean of the results.
The standard deviation of the majority of the results was below $5\%$ of the mean.

Through the tracing facility in the kernel (\code{ftrace}), 
we identified that three benchmarks in the Metis suite use \code{mprotect} extensively.
Those applications are wc (word count), wr (inverted
index calculation) and wrmem, which is a variant of wr that 
allocates a chunk of memory and fills it with random ``words'' instead of  
reading its input from a file. We used default input files for wc and wr,
and 2GB input size for wrmem.
The tracing also revealed that the majority of the calls to \code{mprotect} (over 99\%) succeed
in the speculative path.
We note that in all other Metis benchmarks, which did not call \code{mprotect} as extensively as the 
other three benchmarks mentioned above, the impact of range locks was negligible. 

\figref{fig:metis-tput} shows the runtime results for wc, wr and wrmem (lower is better).
Up to 8--16 threads, all variants perform similarly and scale linearly with the number for threads.
However, once the thread counts increase, and with them the contention on the VM subsystem, 
the variants produce different results.
Notably, the performance of the stock version worsens with the increased contention, while the list-based
range lock variants remain mostly flat, or continue to scale, as in the case of wrmem and \code{list-refined}.
In general, the tree-based range locks perform worse than the list-based ones, and mostly worse even when 
compared with the stock version. 
We believe this is at least in part because of the contention created on the spin lock protecting the access to
the range tree.
Refining the ranges of the range lock acquisitions helps both tree-based and list-based variants, i.e.,
\code{tree-refined} outperforms \code{tree-full}, while \code{list-refined} outperforms \code{list-full}.
In fact, at $144$ threads, \code{list-refined} has 9$\times$ speedup over \code{stock} in wrmem. Again, similar 
to our observation in the user-space skip list experiment, the higher parallelism achieved by list-based 
range locks outweighs the linear complexity of list traversal, even with large number of concurrent ranges (144 in this case). 

It is interesting to note that \code{list-full} outperforms \code{stock} under high contention despite 
always acquiring the range lock for the full range.
We conjecture that this is due to the different waiting policies employed by those two variants.
Specifically, \code{stock} uses a read-write semaphore (\code{mm\_sem}), in which threads block 
(after spinning for a while if optimistic spinning is enabled) when the semaphore is unavailable until they are
waken up by another thread.
In \code{list-full} (and \code{list-refined}), threads block for a small period of time if the range is unavailable and recheck the range, 
which turns to be more efficient under contention.
Exploring different waiting policies and their impact on lock performance is an active area of research~\cite{Dice17, KMK17}.

\figref{fig:metis-tput-refinement} drills down into the effect of refining ranges on the performance of the list-based range locks.
Here \code{list-pf} (\code{list-mprotect}) denotes the variant where only the range in the page fault routine 
(\code{mprotect} operation, respectively) is refined.
As expected, the refinement in the page fault routine does not have much effect, since 
the range lock is acquired there for read while in all other places it is acquired for the full range.
At the same time, refining the range in \code{mprotect} has a small, but positive effect as now \code{mprotect} operations
on non-overlapping ranges can be applied concurrently.
As \figref{fig:metis-tput-refinement} shows, however, it is the combination of the two optimizations that makes a difference --
\code{list-refined}, which refines the range in both page faults and \code{mprotect} and thus allows their concurrent execution, 
substantially outperforms all other variants.

\begin{figure*}[!th]
\subfloat[][wr]{\includegraphics[width=0.33\linewidth]{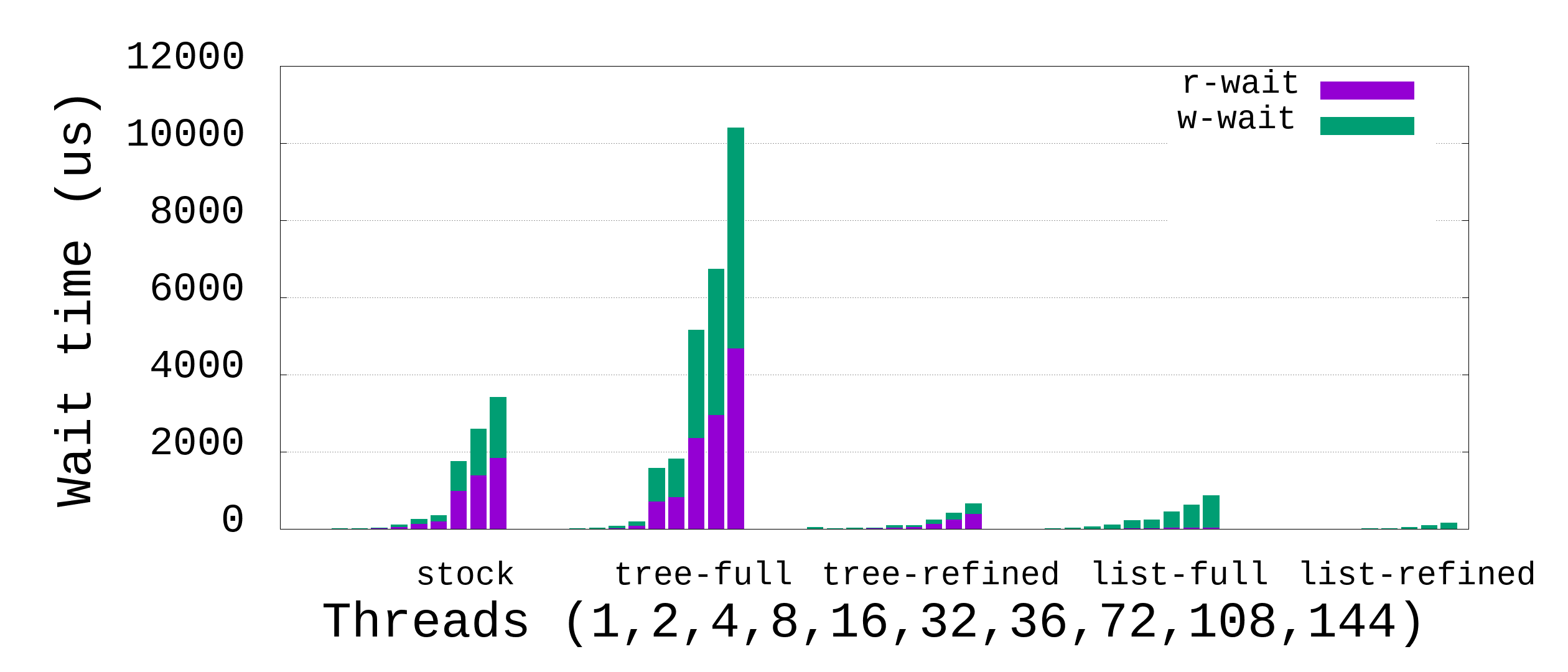}}
\subfloat[][wc]{\includegraphics[width=0.33\linewidth]{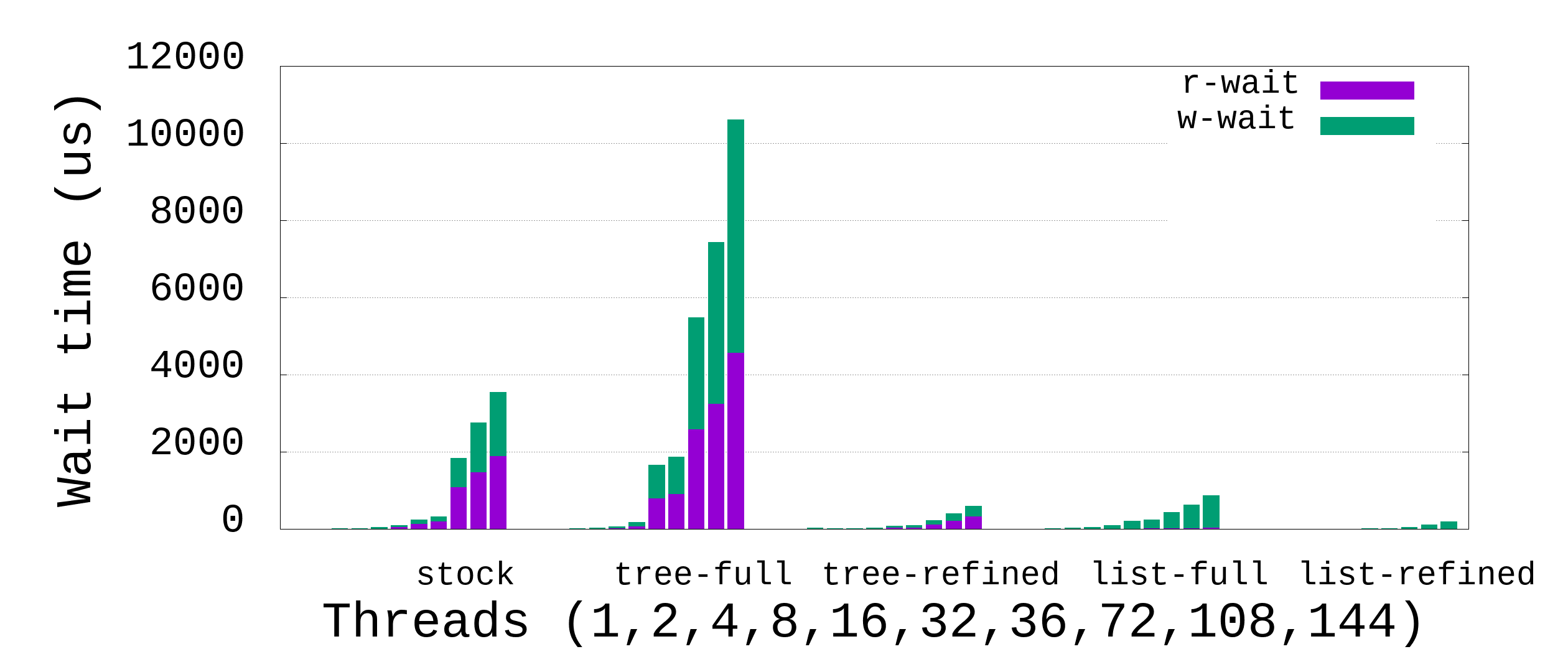}}
\subfloat[][wrmem]{\includegraphics[width=0.33\linewidth]{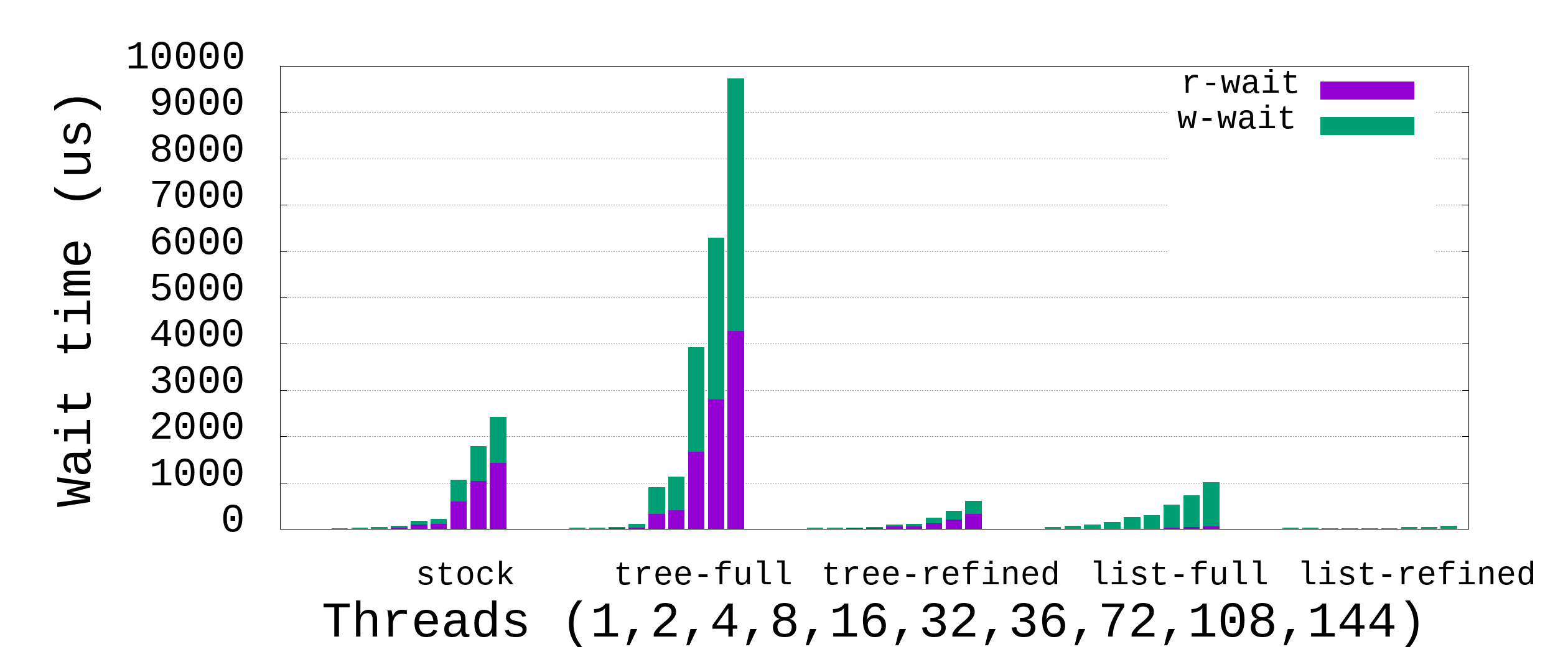}}
\caption{Average wait time for \code{mm\_sem} (in stock) and range lock (in all other variants).}
\figlabel{fig:metis-wait-times}
\vspace{-4mm}
\end{figure*}

\begin{figure*}[!th]
\subfloat[][wr]{\includegraphics[width=0.33\linewidth]{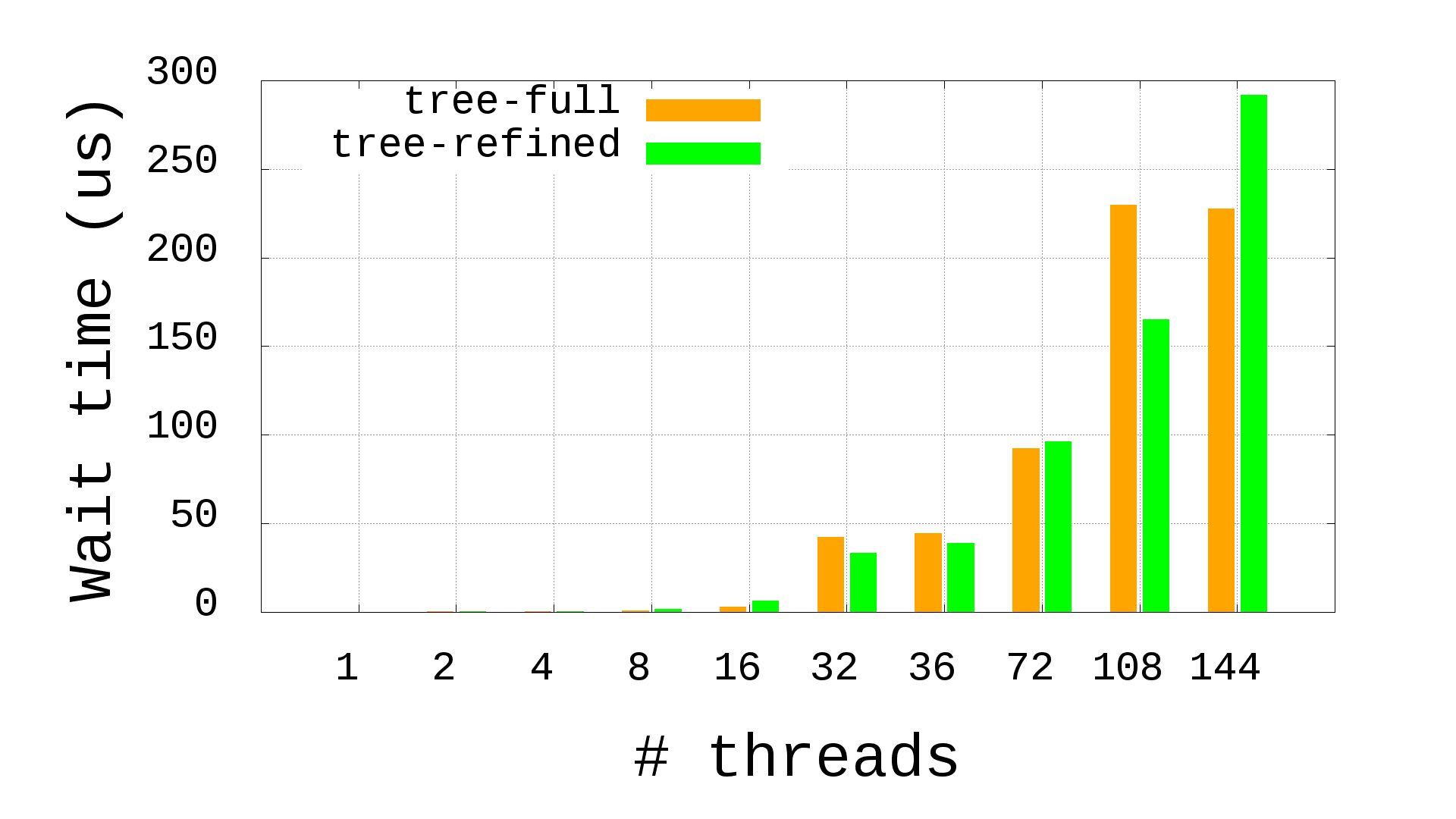}}
\subfloat[][wc]{\includegraphics[width=0.33\linewidth]{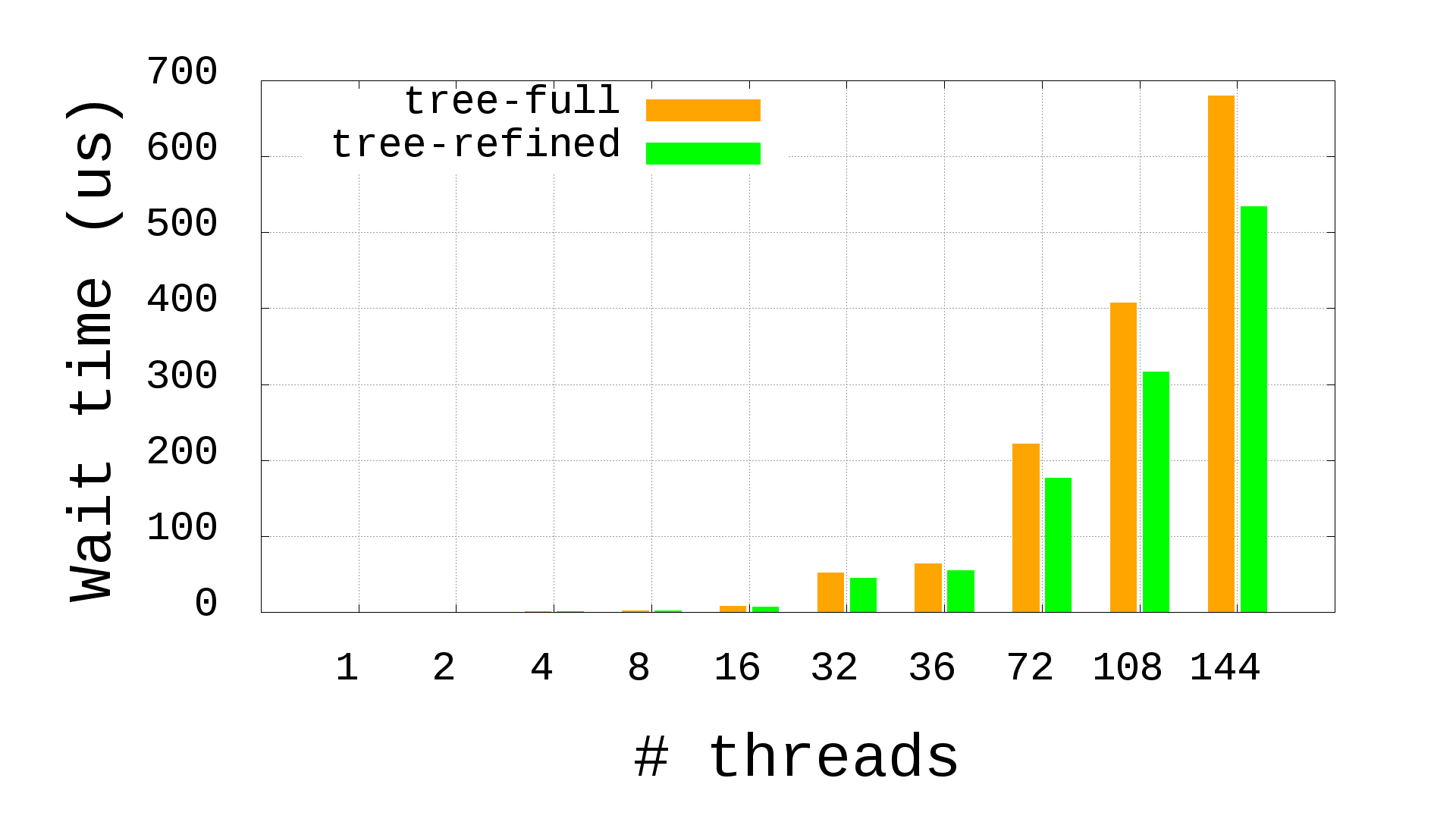}}
\subfloat[][wrmem]{\includegraphics[width=0.33\linewidth]{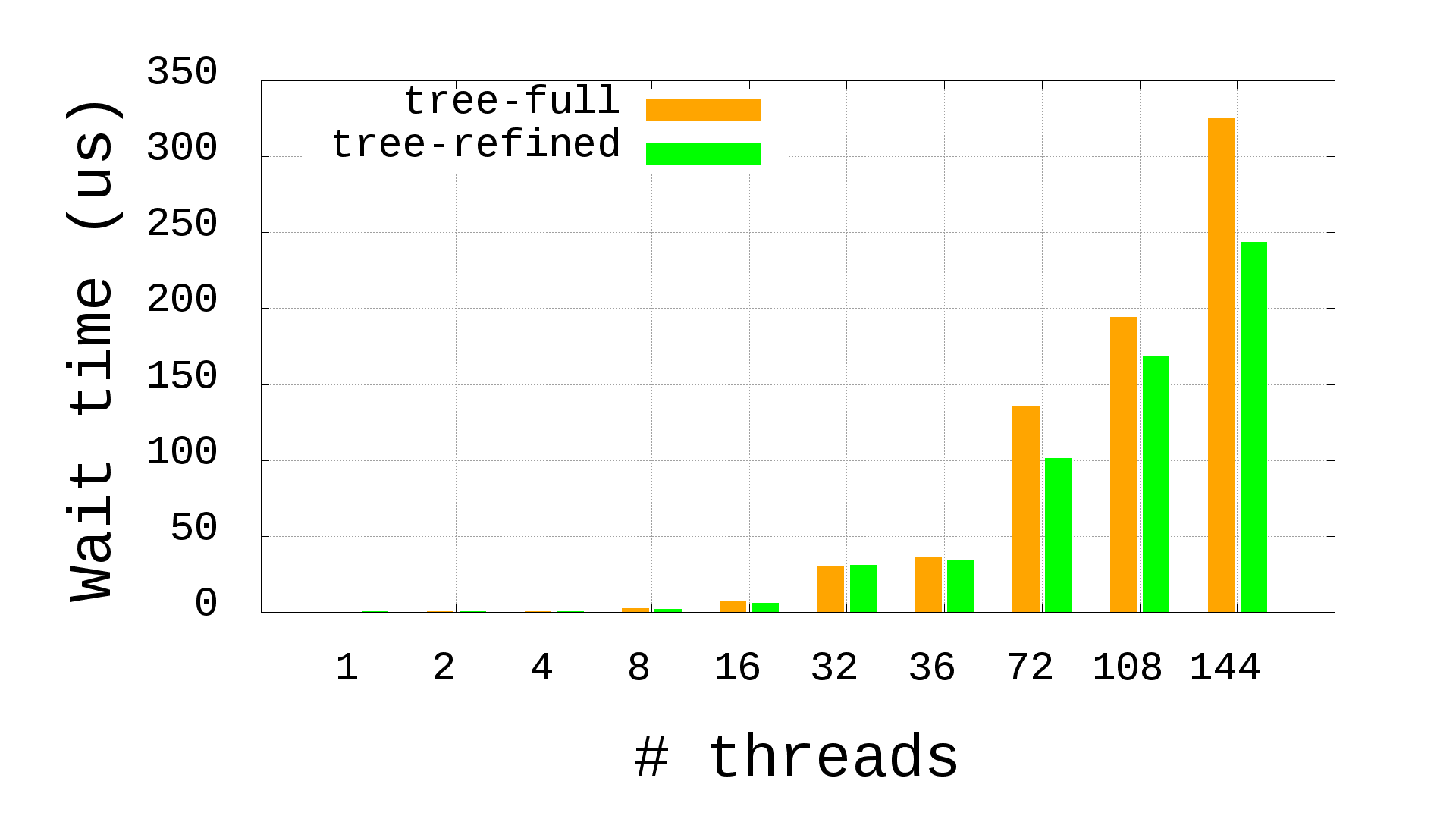}}
\caption{Average wait time for the spin lock protecting the range tree in \code{tree-full} and \code{tree-refined}.}
\figlabel{fig:metis-tree-lock-wait-times}
\end{figure*}

Through the \texttt{lock\_stat} mechanism built into kernel,
we collected statistics on the time threads spent waiting for various locks in the kernel.
(The \texttt{lock\_stat} mechanism is known to introduce a probe effect~\cite{DK19}, therefore it was enabled only for 
runs in which we collected statistics on lock wait times.)
In \figref{fig:metis-wait-times} we plot the average wait times for \code{mm\_sem} (in the stock variant) 
as well as for the range lock in all other variants, breaking down between read and write acquisitions.
Not surprisingly, those results show a (rough) correlation between high wait times and poor scalability.
They also reveal that with range refinement, the average wait times decrease.

\figref{fig:metis-tree-lock-wait-times} shows the average wait time on the spin-lock protecting the range tree in the 
 \code{tree-full} and \code{tree-refined} variants.
Notice that the waiting time grows with the number of threads, supporting our hypothesis that this lock 
represents a point of contention.
The range refinement does not change much the wait time for the spin lock.
This is not surprising, as this lock is acquired for every acquisition of the range lock, regardless of whether or not the range
is available.
However, while in \code{tree-full} the wait time for the spin lock is relatively small compared to the wait time for the range 
lock itself (which includes waiting for a range to become available), in \code{tree-refined} takes the lion share of the range lock wait time 
(cf.~\figref{fig:metis-wait-times} and~\figref{fig:metis-tree-lock-wait-times}).
This underscores the effectiveness of range refinement in allowing parallel processing of the VM operations.
That is, when the ranges are refined, most wait time for a range lock can be attributed to the wait time on the 
auxiliary spin lock rather than waiting for the range availability.
Unlike tree-based range locks, list-based range locks do not have a central point of contention and thus can 
take better advantage of this parallelism, as demonstrated by the results in~\figref{fig:metis-tput}.

\section{Conclusion}
In this paper, we presented the design and implementation of new scalable range locks.
Those locks employ a simple underlying structure (a concurrent linked list) to keep track of acquired ranges.
This structure allows simple lock-less modifications with just one atomic instruction.
Therefore, our design avoids the pitfall of existing range locks, and does not require an auxiliary lock in the common case.

Furthermore, we show how range locks can be employed effectively to 
mitigate the contention on the access to the VM subsystem and
its data structures, in particular, the red-black tree holding VMA structures.
We achieve that through a speculative mechanism introduced into the \code{mprotect} operation;
this mechanism allows to refine the range of the lock acquired in \code{mprotect}.
We also refine the range of lock acquisitions in page fault routines.
Together, those refinements allow parallel processing of page faults and \code{mprotect}s operating on non-overlapping regions of VM space,
which is particularly beneficial, e.g., for the standard GLIBC memory allocator.
In addition, we demonstrate the utility of range locks for the design of concurrent, scalable data structures 
through the example of a range-lock based skip list.

We evaluate the scalability of the new range locks in user-space through several microbenchmarks and kernel-space through several
applications from the Metis suite.
The results show that the new range locks provide superior performance compared to the existing range locks (in the user-space and
kernel), as well as to the current method of VM subsystem synchronization in the kernel (that uses a read-write semaphore).
Future work includes evaluating range locks with additional benchmarks, and 
exploring the usage of range locks in other contexts,
such as parallel file systems~\cite{KKK19} and as building blocks for other concurrent data structures, such as hash tables and binary search trees.

\bibliographystyle{ACM-Reference-Format}
\bibliography{refs}


\end{document}

%% file: defs.tex
\newcommand{\rw}{reader-writer\xspace}
\newcommand{\RW}{Reader-Writer\xspace}

\newcommand{\var}[1]{\lstinline+#1+}
\newcommand{\cSetP}[1]{\lstinline+Set<#1>+}

\newcommand{\figlabel}[1]{\label{figure:#1}}
\newcommand{\nakedfigref}[1]{\ref{figure:#1}}
\newcommand{\figref}[1]{Figure~\nakedfigref{#1}}

\newcommand{\linelabel}[1]{\label{line:#1}}
\newcommand{\nakedlineref}[1]{\ref{line:#1}}
\newcommand{\lineref}[1]{Line~\nakedlineref{#1}}
\newcommand{\linerangeref}[2]{Lines~\nakedlineref{#1}--\nakedlineref{#2}}

\newcommand{\seclabel}[1]{\label{sec:#1}}
\newcommand{\nakedsecref}[1]{\ref{sec:#1}}
\newcommand{\secref}[1]{Section~\nakedsecref{#1}}

\newcommand{\remove}[1] {}
\newcommand{\extabstract}[1] {}
\newcommand{\code}[1] {\texttt{#1}}

%% file: related.tex

\section{Related Work}
\label{sec:related}

Range locks (or byte-range locks) were conceived in the context of file systems to support 
concurrent access to the same file~\cite{unix-manual}.
Since files are a continuous range of bytes, different processes can access 
disjoint regions within the same file if they acquire a (range) lock for 
the desired region, e.g., through the \code{fcntl} operation in Unix~\cite{unix-manual}.
More recently, range locks gained attention as an important piece in the design of 
parallel and distributed file systems. In GPFS~\cite{SH02}, for instance, when a 
process requests access to a region within a file, it is granted a token for the 
whole file. 
Only when another process requests access to another disjoint 
region within the same file, a revoke request is sent to the token holder to 
revoke its rights for the other process's desired range. 
This design has low locking overhead when a file is accessed by a single process 
at the cost of higher overhead when coordination between multiple processes is required. 
Thakur et al.~\cite{Thakur05} suggested the 
use of a data-structure with a per-process entry. 
Each process would acquire a range lock in two steps: first it accesses its slot within the 
data-structure updating it with the desired range and then it reads a snapshot of the data-structure. 
If no other process has requested a conflicting range, the lock is acquired; 
otherwise, the steps are repeated after processes reset their slots within the 
data-structure. 

To avoid liveness issues, Aarestad et al.~\cite{Aarestad06} proposed using a 
red-black tree to store the ranges acquired by different processes.
The same approach is taken by recent 
efforts within the Linux kernel development community to replace the 
read-write semaphore within the virtual memory sub-system with a 
red-black tree-based range lock implementation~\cite{Kar13,Bue17}.
However, as explained earlier, relying on a red-black tree protected by a 
spin lock can be a serious scalability bottleneck, as we will confirm later in 
Section~\ref{sec:eval}.
At the same time, our approach does not use locks in the common case.

In a recent and highly relevant work~\cite{KKK19}, Kim et al. consider using range locks
in the context of parallel file systems, and make a similar observation
regarding the lack of scalability of the existing kernel range locks. 
They followed an alternative design for range locks, which was previously proposed by Quinson et al.~\cite{Quinson09}, in which the entire range is divided into
(a preset number of) segments, each associated with a reader-writer lock. 
To acquire a certain part of the range for read or write, one needs to acquire the reader-writer locks
of the corresponding segments in the respective mode.
In their proposal, the full range acquisition is particularly expensive, as it requires acquiring all 
underlying reader-writer locks.
Moreover, choosing the right granularity, i.e., the number of segments, is critical --- too few segments would
create contention on the underlying reader-writer locks, while too many segments would make range 
acquisition more expensive --- yet, Kim et al. do not discuss how the granularity should be tuned.
Therefore, we believe the applicability of Kim et al.'s scenarios is limited to the cases where the size of the 
entire range and the granularity of the access are known and static, which is precisely the case considered in~\cite{KKK19}.
Nevertheless, we include Kim et al.'s range locks in our performance study in Section~\ref{sec:eval}.

\remove{
To overcome the central synchronization bottleneck incurred by the described solutions, 
Quinson et al.~\cite{Quinson09} proposed a decentralized solution where the 
complete range is divided statically into constant partitions. Each partition is 
protected by a token that must be held by a process before accessing the 
respective range. To reduce the overall communication cost, processes are arranged in a 
tree with the root of the tree being the tail of a distributed queue and processes are allowed 
to communicate only with their parent processes. 
When compared to our solution, such a mechanism can suffer from high latency in lock 
acquisition if the previous lock owner is far apart in the tree. 
Furthermore, the static partitioning of the range can limit 
parallelism due to false sharing when non-overlapping range lock requests fall 
within the same partition.
}

The database community developed a similar concept to range locks, 
known as key-range locks~\cite{Mohan90, Mohan92}.
They were introduced to guarantee serializability of 
database transactions operating on a range of records, 
avoiding so called \emph{phantom read} phenomena~\cite{Lomet93}. 
Besides locking all existing keys within a range, key-range locks also lock the neighboring 
key such that, e.g., no concurrent transaction could insert new keys 
--- that did not exist a priori, and thus could not be locked --- within the 
desired range~\cite{Mohan90}. 
To allow more concurrency, Lomet~\cite{Lomet93} introduced 
hierarchical locking to attribute different lock modes to ranges and keys (e.g., 
locking a range in exclusive mode and a key in shared mode). 
To overcome the high locking overhead incurred by locking all the keys within a range, 
Graefe~\cite{Graefe07} suggested dynamically switching between different locking 
granularities. In addition to the higher locking overhead that these solutions can incur,
they also suffer from lower parallelism since non-overlapping ranges within a region where no 
keys exist have to be unnecessarily serialized on an existing key.
Lomet and Mokbel~\cite{Lomet09} tried to decouple locking from the existing data 
by statically partitioning tables into disjoint partitions. 
Compared to our solution, such an approach suffers
from lower parallelism due to false sharing when 
non-overlapping range lock requests fall within the same partition.

As mentioned earlier, one of the main motivations behind the renewed interest in 
range locks is to design a scalable locking mechanism for the kernel address space operations.
Song et al. attempted to address this problem in the context of parallelizing live VM migration~\cite{SSL13}.
To that end, they proposed a range lock implementation based on a skip list protected by a spin lock.
Conceptually, their design is very similar to the one found in the Linux kernel~\cite{Kar13}.
In particular, both cases have the same bottleneck in the form of a spin lock protecting their corresponding 
underlying data structures for tracking acquired ranges.
 
Several works pursued the same goal of scaling kernel address space operations via a different route: 
replacing the red-black-tree  \code{mm\_rb} with alternative data-structures. 
Clements et al.~\cite{CKZ12} proposed using a RCU-balanced tree to allow 
concurrency between a single writer and multiple readers. In addition to not allowing parallel 
update operations, the proposed tree trades fewer rotations for tree imbalance, which can increase 
tree traversal times.
In another work by the same authors, they proposed using a radix tree, where each mapped page 
will be inserted in a separate node within the tree~\cite{Clements13}. 
Such design supports concurrent read and update accesses to non-overlapping nodes. 
However, this comes at two significant costs: \emph{(i)}~a large memory footprint for using per-page nodes, 
and \emph{(ii)}~high locking overhead, since locking a range of pages entails locking several nodes within the tree.
Unlike both proposals by Clements et al., our work does not require changing \code{mm\_rb}
and thus requires less intrusive changes to the kernel.

%% file: algo.tex

\section{Scalable Range Lock Design}
\subsection{Exclusive Access Variant}
\seclabel{sec:exclusive}

We start with a simpler version of our linked list-based range locks algorithm
intended for mutual exclusion, i.e., it supports concurrent acquisition of disjoint ranges, but no
overlapping ranges are allowed.
In the next section, we describe an extension of the algorithm to support \rw exclusion,
where readers can acquire overlapping ranges, but a writer cannot overlap with 
another (reader or writer) thread.

The idea at the basis of the algorithm is to insert acquired ranges in a linked list sorted by 
ranges' starting points. Accordingly, any overlapping ranges 
will compete to be inserted at the same position in the list. Therefore, 
by relying on an atomic compare-and-swap (CAS) primitive, it is possible to ensure 
that only one range from a group of overlapping ranges will succeed in 
entering the list while others will fail.

\remove{

In this section we describe our list-based range locks algorithm, which 
mitigates the scalability bottleneck of current tree-based 
implementations. This is mainly due to two reasons: \emph{(i)}~we do not rely 
on locks, at least in the common path, to handle a range lock request, and 
\emph{(ii)}~the unlock request is wait-free. The idea at the basis 
of the algorithm is to insert the ranges in a linked list sorted by 
their starting points. Accordingly, any overlapping ranges 
will compete to be inserted at the same position in the list. Therefore, 
by relying on Compare-And-Swap (CAS) primitive it is possible ensure 
that only one rage from a group of overlapping ranges, trying to insert themselves in the list, will succeed in 
entering the list while others will fail.  

For the sake of clarity, we start first by explaining how the list-based 
range locks algorithms works with exclusive-only range locks. Then we 
we describe how the algorithm can be extended to support, both, 
shared and exclusive range locks.
}

\input{algo-mutex-rl.tex}

\sloppy
The pseudo-code for the exclusive access list-based range locks algorithm is shown 
in Listing~\ref{alg:mutex-rl}.
It presents the lock structures and the implementation of the \code{MutexRangeAcquire} and \code{MutexRangeRelease} functions
as well as the auxiliary functions called by those two. 
For the clarity of exposition, we assume sequential consistency. 
Our actual implementation uses \code{volatile} keywords and memory fences where necessarily.
CAS and FAA indicate opcodes for the compare-and-swap and fetch-and-add atomic instructions, 
respectively\footnote{It is easy to simulate FAA with CAS on architectures that do not have a native support for the former.}; \code{Pause()} is a no-op operation used for polite busy-waiting.

For each shared resource protected by a range lock, a \code{ListRL} list must be defined. 
Each node, \code{LNode}, within the list contains the range it defines and a pointer to the next node in the list (cf.~Listing~\ref{alg:mutex-rl}).
At the beginning, the head of the 
list points to \code{null}, indicating that the list is empty. 

When a thread requests an exclusive access over the given region within a resource, 
it first creates an instance of the \code{RangeLock} structure (cf.~\lineref{alg:mutex-rl:acquire-2}), which
contains a pointer to the \code{LNode} structure.
Note that for simplicity, we allocate a new \code{RangeLock} instance each time the \code{MutexRangeAcquire} is called.
It is possible, however, to maintain and reuse a pool of \code{RangeLock} instances; we discuss memory management 
of those instances in detail in \secref{sec:memory}.
Next, the thread initializes the \code{RangeLock} structure  (cf.~\linerangeref{alg:mutex-rl:acquire-3}{alg:mutex-rl:acquire-6}).
Finally, in order to acquire a range, the thread must successfully insert the corresponding node 
into the given range lock list structure (cf.~\lineref{alg:mutex-rl:acquire-7}).
To release the acquired range (in \code{MutexRangeRelease}), a node corresponding to the range is deleted from the list (cf.~\lineref{alg:mutex-rl:release-2}).

The \code{InsertNode} function describes the logic of inserting a node (\code{lock}) into the list (cf.~Listing~\ref{alg:mutex-rl}).
At the high level, this function traverses the list searching for the insertion point (in the increasing order of start addresses) 
for the given node describing the given range.
If the traversal comes by a node with an overlapping range, it waits until that node is removed from the list.

In more detail, \code{InsertNode} traverses the list from its head and checks each node, 
\code{cur}, it encounters while maintaining a pointer, \code{prev}, that points to the address of the previous node's \code{next} pointer. 
A node in the list can either be \textit{marked}, i.e., logically deleted with the least significant bit of its \code{next} pointer being set, or not. 
(We describe the deletion mechanism in detail later.)
If \code{prev} is found to be logically deleted, the traversal has to restart, as the list might have changed in a way that would not
allow the thread to insert its node safely (cf.~\lineref{alg:mutex-rl:marked-prev}).
If \code{cur} is logically deleted, an attempt to remove it 
from the list is made by making \code{prev} point to \code{cur}'s successor (cf.~\linerangeref{alg:mutex-rl:marked:begin}{alg:mutex-rl:marked:end}). 
This is done by issuing CAS to atomically replace the pointer to \code{cur} by a pointer to \code{cur}'s successor. 
Regardless of the result of CAS, which may fail due to a concurrent thread performing the same change,
the traversal of the list continues (\lineref{alg:mutex-rl:marked:end}). We note, however, that in our actual 
implementation we check the result of CAS, and if successful, we reclaim the node using the memory management 
mechanism described in \secref{sec:memory}.

When an unmarked \code{cur} is encountered, the ranges of both \code{cur} 
and \code{lock} are compared (cf.~\lineref{alg:mutex-rl:call-compare}) --- 
see the \code{compare} function for details (\linerangeref{alg:mutex-rl:compare}{alg:mutex-rl:compare-end}). 
If (the range in) \code{lock} succeeds (the range in) \code{cur} without overlapping with it, the list traversal is continued 
(\linerangeref{alg:mutex-rl:succeeds:begin}{alg:mutex-rl:succeeds:end}). 
If they overlap, then \code{lock} must wait until \code{cur} is marked as deleted,
which will happen when the thread 
that acquired the corresponding range exits its critical section.
After the wait, the traversal resumes from the same point (and the marked \code{cur} 
will be subsequently removed as described above).

In case \code{lock} precedes \code{cur} (or \code{cur} is \code{null}), 
the insertion position for \code{lock} has been found to be between \code{prev} and \code{cur}.
To execute the insertion, CAS is issued trying to replace \code{cur} by \code{lock} in \code{prev}
(see \lineref{alg:mutex-rl:insert1}). 
If the CAS is successful, the exclusive access over the range is now acquired and the function 
can return (\lineref{alg:mutex-rl:success}).
Otherwise, this means another 
thread has changed \code{prev}, either by inserting a node right after \code{prev} or marking \code{prev} for deletion.
In this case, the traversal is resumed from the same point with \code{cur} being updated to a new value from \code{prev} (\lineref{alg:mutex-rl:resume}).

An acquired range lock is unlocked by deleting the corresponding node from the list (cf.~\lineref{alg:mutex-rl:release-2}).
In a linked list, it means updating the \code{next} pointer of the node's predecessor to point to the node's successor.
However, one has to locate the predecessor first, which means traverse the list from the head.
For performance considerations, when releasing a range lock, we only delete the corresponding node logically.
This is achieved by a common technique in concurrent linked list implementations of \emph{marking} the node~\cite{Harris01}, 
i.e., setting the least significant bit (LSB) of its \code{next} pointer.
This setting avoids races with concurrent threads trying to change 
the value of \code{next} while inserting or removing a neighboring node.
(Recall that CAS instructions in \code{InsertNode} are issued on pointers of nodes that are expected not to be logically deleted.)
Since only one thread can mark any given node (the thread that acquired the corresponding range), 
setting the LSB can be done with an atomic increment instruction (cf.~\lineref{alg:mutex-rl:mark}).
This means that on architectures that support such an instruction, range lock release is wait-free.
As described above, marked nodes are removed from the list during traversals in \code{InsertNode}.

\noindent \textbf{Correctness Argument:}
We argue that the pseudo-code in Listing~\ref{alg:mutex-rl} is a correct and a deadlock-free implementation of exclusive access range locks.
For correctness, we argue that the implementation never allows two threads to acquire range locks with overlapping ranges.
This claim is based on the following invariant:

\begin{invariant}
For any two consecutive ranges R1 and R2 in the list ListRL, R1.end $\leq$ R2.start.
\end{invariant}


To prove the progress property, we note that a thread $T$ would remain infinitely long in the \code{InsertNode}
function only if (a) it finds infinitely often its \code{prev} variable pointing to a deleted node (cf.~\lineref{alg:mutex-rl:marked-prev}),
or (b) it traverses infinitely many logically deleted nodes (cf.~\linerangeref{alg:mutex-rl:marked:begin}{alg:mutex-rl:marked:end}),
or (c) it traverses infinitely many ranges that end before the thread's range starts (cf.~\linerangeref{alg:mutex-rl:succeeds:begin}{alg:mutex-rl:succeeds:end}),
or (d) it waits infinitely long to a thread with an overlapping range (cf.~\lineref{alg:mutex-rl:overlaps:end}).
Given that the list contains a finite number of nodes when $T$ calls  \code{InsertNode},
cases (a), (b), and (c) are possible only if some other thread (or threads) insert
(and delete) infinitely many nodes, which in turn means that those threads acquire and release 
infinitely many ranges while $T$ is executing \code{InsertNode}.
Assuming that no thread fails while holding the range lock, then either case (d) is impossible as the thread would mark its node as logically deleted in a finite number of steps (if the hardware supports wait-free FAA), or case (d) is possible only if infinitely many threads would acquire and release a range lock (for the CAS-based implementation of FAA).
Thus, $T$ would either return from \code{InsertNode} (and thus acquire the range lock), or 
infinitely many threads would acquire and release the range lock while T is executing \code{InsertNode}.

We note that the described implementation of the list-based range lock is not starvation-free, e.g., a thread trying to insert a node into the
list may continuously fail to apply CAS (cf.~\lineref{alg:mutex-rl:insert1}) and/or be forced to restart the traversal if its \code{prev} 
pointer gets marked (cf.~\linerangeref{alg:mutex-rl:marked-prev}{alg:mutex-rl:marked-prev-end}).
In \secref{sec:fairness} we describe a simple mechanism to introduce fairness and avoid starvation.

\subsection{\RW Variant}

In the previous section, we have presented a range lock algorithm that supports 
acquiring exclusive access on defined ranges. 
Now, we extend the algorithm to handle reader-writer synchronization.
For the sake of brevity, in this section threads acquiring a range lock in shared mode will be referred to as readers 
while threads acquiring a range lock in exclusive mode will be referred to as writers.

A natural way to extend the range locks algorithm from the previous section
is to consider the access mode (read or write) in the \code{compare} function,
and allow an overlap when both compared ranges belong to readers.
In other words, we would traverse the list (in \code{InsertNode}) and insert the given 
node into the list even if that node (i.e., its range) overlaps with an existing node, and both nodes belong to readers.

\begin{figure}[t]
\centering
\subfloat[][]{
\begin{adjustbox}{minipage=\linewidth,scale=0.5}
\includegraphics[width=1\linewidth, Clip=0 5cm 0cm 0cm]{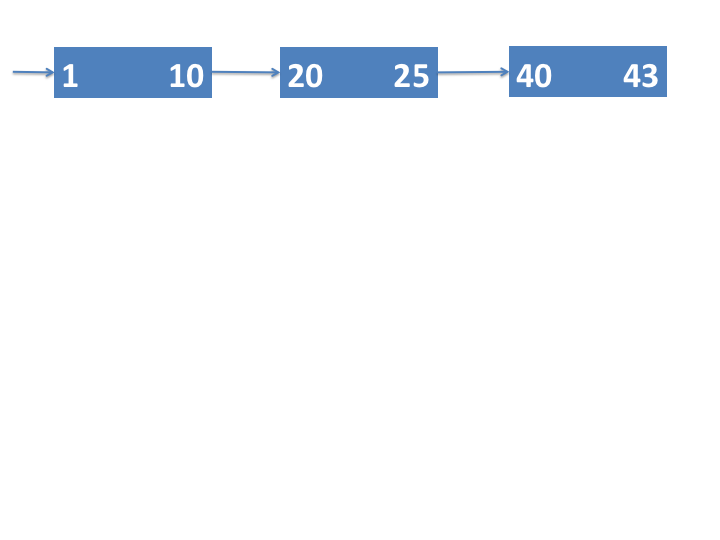}
\end{adjustbox}
}
\subfloat[][]{
\begin{adjustbox}{minipage=\linewidth,scale=0.5}
\includegraphics[width=1\linewidth, Clip=0 3cm 0cm 0cm]{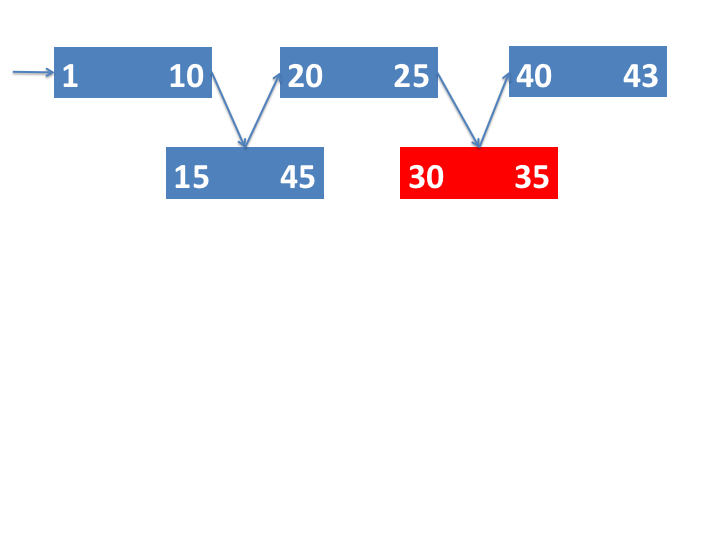}
\end{adjustbox}
}
\caption{An example for a race condition between readers and writers solved by validation.
(a): Three reader ranges are in the list.
(b): A new reader with the range \code{[15..45]} arrives, and since it starts before the reader with the range \code{[20..25]}, it inserts itself into the list after a reader with the range \code{[1..10]}. At the same time, a writer with the range \code{[30..35]} arrives, finds that it does not overlap with any reader and inserts itself into the list after the reader with the range \code{[20..25]}.}
\figlabel{fig:example}
\end{figure}

Unfortunately, this approach enables a race condition between readers and writers, exemplified in \figref{fig:example}.
A reader may ``miss'' a writer with an overlapping range located down the list.
At the same time, a writer may ``miss'' a reader with an overlapping range that entered the list at the point that the writer has already traversed.
This race condition is possible because overlapping readers and writers may insert themselves into the list at 
different points (i.e., after different nodes), and therefore they do not compete to modify the same (\code{next}) pointer (see \figref{fig:example}).

We solve this problem with an extra validation step performed by readers and writers.
Specifically, when a reader inserts its node into the list, it continues to scan the list until it finds a node 
with a range that does \emph{not} overlap.
If during this scan the reader comes across a writer, it waits until the writer's node is (logically) deleted.
As for the writer, its validation step is slightly different (since
a similar wait by writers for readers would lead to deadlock).
Once the writer inserts itself into the list, it re-traverses the list from the head until it finds its own node.
If during this re-traversal, a writer finds a reader with an overlapping range, 
the writer leaves the list (by logically deleting its node) and restarts the acquisition attempt from the beginning.
The race conditions can only happen between a reader and a writer that have both inserted 
themselves into the list, therefore re-traversing the list will guarantee detecting such a race. 

Note that this validation approach may cause starvation of writers, as they may be forced to restart repeatedly by incoming readers.
We describe a way to avoid this issue in \secref{sec:fairness}.
Furthermore, note that our validation approach gives preference to readers, since in case of a conflict they stay in the list while writers restart.
It is straightforward to reverse the scheme and give preference to writers instead, by letting them stay in the list 
(while waiting for conflicting readers to leave) and making the readers restart in case of a conflict.

 
\input{algo-rw-rl.tex}

Listing~\ref{alg:rw-rl} shows how to implement shared range 
locks, where overlapping acquisitions with shared (reader) accesses do not 
block each the other.
The pseudo-code in Listing~\ref{alg:rw-rl} is presented in the form of diffs from Listing~\ref{alg:mutex-rl}.
The \code{LNode} structure includes now a flag (\code{reader}) indicating whether 
the corresponding range is acquired for read or for write. (This is a trivial change and thus not shown).
The \code{RWRangeAcquire} function is similar to \code{MutexRangeAcquire} (in Listing~\ref{alg:mutex-rl}),
except that the call to \code{InsertNode} is now wrapped in a \code{do-while} loop.
This loop will be executed more than once by a writer only, and only in the case the writer's validation fails.
The \code{RWRangeRelease} function is identical to \code{MutexRangeRelease} (in Listing~\ref{alg:mutex-rl}) and thus not shown.
The \code{compare} function is adapted in a straightforward way to allow overlapping reader ranges (see \linerangeref{alg:rw-rl:compare}{alg:rw-rl:compare-end}).
Finally, the only change in the \code{InsertNode} function is the call to validation functions 
according to the access mode for which the range lock is acquired (see \linerangeref{alg:rw-rl:insert1}{alg:rw-rl:insert2}).

The details of the validation functions are given in Listing~\ref{alg:rw-rl-aux}.

A reader executes the \code{r\_validate} function, where it continues to traverse the list from the point where 
it just inserted its node and until it either reaches the end of the list
or reaches a node that starts after the reader's node ends (\lineref{alg:rw-rl:rvalidate-5}).
During the traversal, and as an optimization, the reader attempts to remove logically deleted nodes from the list (\linerangeref{alg:rw-rl:rvalidate-6}{alg:rw-rl:rvalidate-9}).
Like mentioned before in \secref{sec:exclusive}, in the actual implementation, successfully  removed nodes 
are recycled using the memory management mechanism described in \secref{sec:memory}.
Furthermore, if it encounters a writer's node, it waits until the node is logically deleted (\linerangeref{alg:rw-rl:rvalidate-13}{alg:rw-rl:rvalidate-15}).

A writer, for its part, executes the \code{w\_validate} function, where it traverses the list from the head until 
it reaches its node (\lineref{alg:rw-rl:wvalidate-5}).
Like a reader, during the traversal the writer attempts to remove logically deleted nodes from the list (\linerangeref{alg:rw-rl:wvalidate-6}{alg:rw-rl:wvalidate-9}).
If, however, a writer comes across an overlapping node, it deletes its node and fails the validation (\linerangeref{alg:rw-rl:wvalidate-13}{alg:rw-rl:wvalidate-15}).
Note that this overlapping node has to belong to a reader, since a writer waits for any overlapping node 
(for which \code{compare} returns zero) before 
inserting itself into the list (cf.~\linerangeref{alg:mutex-rl:overlaps:begin}{alg:mutex-rl:overlaps:end} in Listing~\ref{alg:mutex-rl}).

\noindent \textbf{Correctness Argument:} 
We argue that the pseudo-code in Listing~\ref{alg:rw-rl} is a correct implementation of  \rw range locks.
To that end, we argue that the implementation never allows two threads to acquire conflicting ranges ---  
ranges conflict when they overlap and at least one of them is a writer. 
Our claim is based on the following invariant:
\begin{invariant}
\label{lem:rw:writer}
For any two consecutive ranges $R1$ and $R2$ in $ListRL$, $R1.start \leq R2.start$. 
Moreover, if $R1$ is a writer, then $R1.end \leq R2.start$.
\end{invariant}

\sloppy
Based on this invariant, if a reader or a writer $G$  in $ListRL$
overlaps with a writer $W$, then $G.start \leq W.start$ (if $W.start \leq G.start$ then, 
according to Invariant~\ref{lem:rw:writer}, $W.end \leq G.start$, thus they can not overlap).  
Assume there is a writer $G$  in $ListRL$ that overlaps with $W$ and $G.start \leq W.start$. 
Since $G$ is a writer then $G.end \leq W.start$ (otherwise Invariant~\ref{lem:rw:writer} breaks), 
a contradiction. 
Now, we are left with the case of $G$ being a reader. 
There are two possibilities: either \emph{(i)}~$G$ entered $ListRL$ before $W$ 
or \emph{(ii)}~after $W$. Note that a range enters $ListRL$ after a successful CAS at 
\lineref{alg:rw-rl:insert1} in Listing~\ref{alg:rw-rl}.

The intuition at the basis of our correctness argument is that if a conflicting range that enters $ListRL$ last 
(among the two conflicting ranges) defers to the other conflicting range, we can guarantee \rw exclusion.  
Accordingly, to handle the first case, \code{w\_validate} is 
executed after the CAS operation at \lineref{alg:rw-rl:insert1}, and since it starts traversing ranges in 
$ListRL$ from the head node, then any range $G$ with $G.start \leq W.start \leq G.end$ that entered $ListRL$ before  
$W$ (and has not left yet) is guaranteed to be visited during the traversal. For the second case, \code{r\_validate} is 
executed after the CAS operation, and since it starts traversing ranges in $ListRL$ from the node succeeding $G$, 
any range $W$ with $G.start \leq W.start \leq G.end$ that entered $ListRL$ before  
$G$ is guaranteed to be visited during the traversal. 
Consequently, by ensuring that both \code{w\_validate} and \code{r\_validate} do not return 
successfully if a conflicting range lock is visited, \rw exclusion is guaranteed.

\sloppy
As for deadlock freedom, the same arguments used for the basic mutual exclusion apply also for the \rw 
pseudo-code. There are two additional 
cases, though, in which thread T may wait infinitely long in \code{InsertNode}: 
(a)~when \code{w\_validate} infinitely often returns 1 and (b)~when \code{r\_validate} function  waits infinitely long for a thread with an overlapping writer range. 
Case (a) is possible if other thread (or threads) insert (and delete) infinitely many overlapping nodes, which in turn means that those threads acquire and release infinitely many reader ranges while T is executing \code{InsertNode}.
Assuming that no thread fails after executing the CAS operation 
at \lineref{alg:rw-rl:insert1}, case (b) is similar to case (d) in the exclusive access variant (see Section \ref{sec:exclusive}).

Similarly to what we mentioned earlier, we note that while the presented \rw range locks are deadlock-free, they are not starvation-free. 
We discuss next how our design can be augmented with an auxiliary lock to avoid starvation.

\subsection{Fairness}
\seclabel{sec:fairness}

The range lock design presented so far does not use any locks.
However, it allows starvation of a thread repeatedly failing to insert its node into the list due to other 
threads concurrently acquiring and releasing locks (and thus modifying the list).
A simple way to avoid that is to introduce an auxiliary (fair) \rw lock coupled with an \code{impatient} counter.
A thread acquiring the range lock checks the \code{impatient} counter, and if it is equal to zero (common case), 
proceeds with the range acquisition.
Otherwise, if the counter is non-zero, it acquires the RW-lock for read.
When a thread fails to acquire the range lock in a few attempts, it bumps up the \code{impatient} counter (atomically) and acquires the RW-lock for write.
The counter is decremented (atomically) upon the release of the RW-lock that was acquired for write.
Note that any race between a thread reading zero from the counter and a thread incrementing the counter is benign,
as the sole purpose of this counter is to introduce fairness rather than ensure the correctness of the underlying range lock. 

\subsection{Memory Reclamation}
\seclabel{sec:memory}

In the proposed design of range locks, threads traverse list nodes concurrently with threads modifying the list.
While this approach avoids the bottleneck of an auxiliary lock protecting the underlying structure as 
found in the existing implementation of range locks~\cite{Kar13, Bue17}, 
the lock-less traversal of a list poses a challenge with respect to the memory management of list nodes.
This is because a list node may not be immediately reclaimed once it is removed from the list, since other threads
traversing the list may have a reference to this node and may try to access its memory after it has been removed from the list.
This is a well-known problem in the area of concurrent data structures~\cite{Michael04, FA17}, and multiple solutions are available~\cite{HMBW07}.

For our kernel-space implementation, we employ the read-copy-update (RCU) method~\cite{MS98}, which is readily supported in the Linux kernel~\cite{MBW12}. 
RCU is a synchronization mechanism that allows readers (threads that access 
shared data without modifying it) to execute concurrently with a writer (a thread modifying shared data) without 
 acquiring locks. The idea at the basis of RCU is for readers to announce when they start and 
finish accessing shared data, while writers apply their changes to a copy of the data that is visible to only 
new readers (i.e., readers that started after the writer). The old data is then atomically replaced by the (modified) copy and recycled when there are no more active old readers. In the context of memory reclamation, threads traversing the list mark 
themselves as readers throughout the traversal, while a thread trying to reclaim memory, performs that operation as a 
writer. 
To facilitate progress and efficiency, we employ the \code{call\_rcu()} API, which does not require
waiting for concurrent readers when retiring memory. That is, the memory will be retired (through the callback passed to \code{call\_rcu()}) asynchronously, after those readers exit their corresponding critical sections. 

For the user-space implementation, we chose an epoch-based reclamation scheme~\cite{Fra04} for its simplicity and low overhead.
We augment the epoch-based reclamation scheme with thread-local object (node) pools to amortize reclamation costs as we detail next.
Each thread maintains two (thread-local) pools of list nodes (where each pool is implemented as a sequential linked list).
One pool contains list nodes ready to be allocated and used for a range lock acquisition (we call this pool \emph{active}), while
another pool contains list nodes that this thread has removed from the list, but has not recycled yet (we call this pool \emph{reclaimed}).
Note that each thread has only two pools, regardless of the number of range locks it accesses.
To amortize allocation costs, the active pools are initialized with $N$ records ($N=128$ in our case), while the reclaimed pool are initially empty.
In addition, each thread is associated with an epoch number, which is a $64$-bit counter initialized to zero and
incremented before (and after) a thread makes first (last, respectively) reference to a list node when traversing the list 
during the range lock acquisition.

When a thread removes a node from the list, it puts the node into the reclaimed pool.
When a thread needs to allocate a new node for the range lock acquisition, it grabs a node from the active pool.
If the active pool is empty, it calls a barrier function, which iterates over epoch numbers of other threads and waits for each thread
to finish its current operation (by incrementing its epoch), if such operation is in progress (i.e., if the corresponding epoch number is odd).
After the barrier, it is safe to recycle (or reclaim) all nodes in the reclaimed pool.
Therefore, the thread switches between its pools, and the (now empty) active pool becomes the reclaimed pool, 
and the (potentially, non-empty) reclaimed pool becomes the active pool.
After the switch, and in order to keep the memory footprint of the system steady, the thread checks whether the size
of the active pool is too small (e.g., has less than $N/2$ nodes) and if so, replenishes the pool by allocating new nodes
(up to the total size of $N$).
At the same time, if the active pool is too large (e.g., has more than $2N$ nodes), the active pool is trimmed by reclaiming (freeing) 
extra nodes (up to the total size of $N$).
Note that when the workload is balanced, i.e., each thread removes roughly the same number of nodes that it inserts into 
the list underlying the range lock, the memory management does not involve the system memory allocator (except for the initial allocation of active pools).

\subsection{Fast Path Optimization}

The proposed range lock implementation is amendable to a fast path optimization, which allows the range 
lock to be acquired and released in a constant number of steps when the lock is not contended.
This is particularly important for a single thread execution, but is also useful when the lock is accessed by multiple threads 
while only one of them accesses the lock at a time.

The fast path is implemented as following.
When a thread acquires the range lock, it checks whether the list is empty 
(i.e., whether \code{head} points to \code{null}).
If so, it attempts to set (using CAS) the \code{head} of the list to the marked pointer to the node 
corresponding to the range lock acquisition request.
If successful, the range lock acquisition is complete.
In pseudo-code, the fast range lock acquisition path is implemented with the following two lines inserted right before the call to 
\code{InsertNode} in the range lock acquisition function (e.g., before \lineref{alg:mutex-rl:acquire-7} in Listing~\ref{alg:mutex-rl}):
\begin{lstlisting}[numbers=none, language=Python, label=alg:mutex-rl-fast-path, escapechar=|]
if (listrl->head == NULL and CAS(&listrl->head, NULL, mark(rl->node))):
  return rl;	
\end{lstlisting}

The (not shown) \code{mark} macro simply sets the LSB of the given pointer.
Note that the \code{head} pointer can be marked only if the lock has been acquired on the fast path.
We exploit this fact in two places.
First, during unlock, if a thread $t$ finds that the \code{head} is marked and points to $t$'s node, $t$ realizes that it has
acquired the range lock through the fast path, and attempts to release it by setting \code{head} to \code{null} (using CAS).
At the same time, if another thread $t^\prime$ attempts to acquire the range lock on the regular path and finds \code{head}
being marked, it first removes the mark (by changing \code{head} to point to the same node but without mark using CAS),
and then proceeds with the acquisition.
This ensures that a range lock $l$ acquired on the fast path would be properly released on the regular path if other threads
acquired other ranges in the meantime between $l$'s acquisition and release.

In summary, the main difference between the fast and regular paths is in the way nodes are removed from the list.
While on the regular path, the node is marked during lock release, and removed during lock acquisition when (possibly) another thread
traverses the list, on the fast path the removal is eager.
This reduces the total number of atomic operations required to delete a node from the list, and keeps the number 
of steps performed during the lock operation constant as there are no marked nodes that are needed to be removed from the list first.

%% file: algo-mutex-rl.tex

\begin{figure}[ht!]
\begin{lstlisting}[language=Python, caption=Pseudo-code for the exclusive access range locks implementation., label=alg:mutex-rl, escapechar=|,deletendkeywords={next}, commentstyle=\color{blue} ]
class LNode: |\hfill| ##defines a node a within the list
	__u64 start; __u64 end
	LNode* next

##defines a list for range locks protecting the same resource
class ListRL: |\hfill| 
	LNode* head  |\hfill|	## pointer to the head of the list

##defines a range lock to protect a region within a shared resource
class RangeLock:
	LNode*	node	

def MutexRangeAcquire(ListRL* listrl, __u64 start, __u64 end): | \linelabel{alg:mutex-rl:acquire-1} |
	RangeLock* rl = new RangeLock()			| \linelabel{alg:mutex-rl:acquire-2} |
	rl->node = new LNode()					| \linelabel{alg:mutex-rl:acquire-3} |
	rl->node->start = start; rl->node->end = end; rl->node->next = NULL| \linelabel{alg:mutex-rl:acquire-6} |
	InsertNode(listtl, rl->node)					| \linelabel{alg:mutex-rl:acquire-7} |
	return rl								| \linelabel{alg:mutex-rl:acquire-8} |
	
def MutexRangeRelease(RangeLock* rl)			| \linelabel{alg:mutex-rl:release-1} |
	DeleteNode(rl->node)					| \linelabel{alg:mutex-rl:release-2} |	

def compare(LNode* lock1, LNode* lock2): | \linelabel{alg:mutex-rl:compare} |
	if !lock1: return 1  |\hfill|##	lock1 is end of the list, no overlap
	## check if lock1 comes after lock2, no overlap
	if lock1->start >= lock2->end: return 1 
	## check if lock1 is before lock2, no overlap 
	if lock2->start >= lock1->end: return -1
  return 0  |\hfill|##	lock1 and lock2 overlap|\linelabel{alg:mutex-rl:compare-end} |

def marked(LNode *node):	return is_odd((__u64)node)
def unmark(LNode *node):	return (__u64)node - 1

def InsertNode(ListRL *listrl, LNode *lock): |\linelabel{alg:mutex-rl:insert}|
	while true:
		LNode** prev = &listrl->head
		LNode* cur = *prev
		while true:
			if marked(cur): |\hfill|## prev is logically deleted? |\linelabel{alg:mutex-rl:marked-prev}|
				break |\hfill|## traversal must restart as pointer to previous is lost.|\linelabel{alg:mutex-rl:marked-prev-end}|
			elif cur and marked(cur->next):|\linelabel{alg:mutex-rl:marked:begin}||\hfill|## cur is logically deleted? 
				LNode *next = unmark(cur->next)
				CAS(prev, cur, next) |\hfill|## try to remove it from list
				cur = next |\hfill|## and continue traversing the list|\linelabel{alg:mutex-rl:marked:end}|
			else: |\hfill|## cur is currently protecting a range
				auto ret = compare(cur, lock) | \linelabel{alg:mutex-rl:call-compare} | 
				if ret == -1: |\hfill|## lock succeeds cur: | \linelabel{alg:mutex-rl:succeeds:begin} | 
					prev = &cur->next |\hfill|## continue traversing ...
					cur = *prev |\hfill|## the list.|\linelabel{alg:mutex-rl:succeeds:end} |
				elif ret == 0: |\hfill|## lock overlaps with cur: |\linelabel{alg:mutex-rl:overlaps:begin}|
					while(!marked(cur->next)): |\hfill|## wait until ...  
						Pause() |\hfill|## cur marks itself as deleted|\linelabel{alg:mutex-rl:overlaps:end}|
				elif ret == 1: |\hfill|## lock precedes cur or reached end of list:
					lock->next = cur |\hfill|## then try to ...
					if CAS(prev, cur, lock): ## insert lock into the list|\linelabel{alg:mutex-rl:insert1}|
						return 0 |\hfill|## success - the range is acquired now. | \linelabel{alg:mutex-rl:success}|
					cur = *prev |\hfill|## o/w continue traversing the list.|\linelabel{alg:mutex-rl:resume} |

def DeleteNode(LNode *lock):
	FAA(&lock->next, 1) |\hfill|## logically mark lock as deleted.|\linelabel{alg:mutex-rl:mark}|
\end{lstlisting}
\end{figure}

%% file: algo-rw-rl.tex

\begin{figure}[tp]
\begin{lstlisting}[language=Python, caption=Reader-Writer range locks presented as diffs from the corresponding functions in Listing~\ref{alg:mutex-rl}., label=alg:rw-rl, escapechar=|,deletendkeywords={next}, commentstyle=\color{blue}]
##function called to protect a given range 
##within a resource protected by list
def RWRangeAcquire(ListRL* listrl, __u64 start, __u64 end, int reader): 
	do:
		RangeLock* rl = new RangeLock()
		rl->node = new LNode()
		rl->node->start = start
		rl->node->end = end
		rl->node->next = NULL
		rl->node->reader = reader |\hfill|## set to 1 if reader, 0 if writer
	while(InsertNode(listrl, rl->node))
	return rl
					
## return values:
## -1: if lock1 comes before lock2, or 
##      if both are readers and lock1 starts before lock2
## 0:  if they overlap (and at least one of the locks is a writer)
## +1 if lock1 comes after lock2, or
##      if both are readers and lock1 starts after lock2
def compare(LNode* lock1, LNode* lock2): | \linelabel{alg:rw-rl:compare} |
	if !lock1: return 1
	int readers = lock->reader + lock2->reader
	if lock2->start >= lock1->end: return -1
	if lock2->start >= lock1->start and readers == 2: return -1
	if lock1->start >= lock2->end: return 1
	if lock1->start >= lock2->start and readers == 2: return 1
  return 0							 | \linelabel{alg:rw-rl:compare-end} |

def InsertNode(ListRL *listrl, LNode *lock): |\label{alg:rw-rl:insert}|
	... |\hfill|## same as InsertNode in Listing |\ref{alg:mutex-rl}| up to| \lineref{alg:mutex-rl:insert1}|
		if CAS(prev, cur, lock):  |\hfill|## try to insert lock into the list|\linelabel{alg:rw-rl:insert1}|
			if lock->reader:	return r_validate(lock) |\hfill|## validate as a reader
			else: return w_validate(listrl, lock) |\hfill|## validate as a writer|\linelabel{alg:rw-rl:insert2}|
	... |\hfill|## same as Listing |\ref{alg:mutex-rl} |
				
\end{lstlisting}
\end{figure}

\ContinueLineNumber
\begin{figure}[tp]
\begin{lstlisting}[language=Python, caption=Validation functions called from \code{InsertNode} in Listing~\ref{alg:rw-rl}., label=alg:rw-rl-aux, escapechar=|,deletendkeywords={next}, commentstyle=\color{blue}]	
## reader validation: scans the list from the its node until 
## a non-overlapping node is found. 
## If an overlapping writer is found, waits for it and resumes the scan.
def r_validate(LNode *lock): | \label{alg:rw-rl:r_validate:begin}|		
		LNode** prev = &lock->next
		LNode* cur = unmark(*prev)
		while true:
			if !cur or cur->start > lock->end: return 0	|\hfill | ## validation is over|\linelabel{alg:rw-rl:rvalidate-5}|
			if marked(cur->next): |\hfill|## cur is logically deleted?|\linelabel{alg:rw-rl:rvalidate-6}|
				LNode *next = unmark(cur->next)
				CAS(prev, cur, next) |\hfill|## try to remove it from list
				cur = next |\hfill|## and continue traversing the list|\linelabel{alg:rw-rl:rvalidate-9}|
			elif cur->reader: |\hfill|## another overlapping reader|\linelabel{alg:rw-rl:rvalidate-10}|
				prev = &cur->next |\hfill|## continue traversing ...|\linelabel{alg:rw-rl:rvalidate-11}|
				cur = unmark(*prev) |\hfill|## the list.|					\linelabel{alg:rw-rl:rvalidate-12}|
			else: |\hfill|## lock overlaps with cur and cur is a writer:|\linelabel{alg:rw-rl:rvalidate-13}|
				while !marked(cur->next): |\hfill|## wait until ...|\linelabel{alg:rw-rl:rvalidate-14}|
					Pause() |\hfill|## cur marks itself as deleted|\linelabel{alg:rw-rl:rvalidate-15}|

## writer validation: scans the list from the head until it finds itself.
## If an overlapping reader exists, deletes itself and validation fails.
def w_validate(ListRL *listrl, LNode *lock):  | \label{alg:rw-rl:w_validate:begin}|
		LNode** prev = &listrl->head
		LNode* cur = unmark(*prev)
		while true:
			if cur == lock: return 0	|\hfill | ## validation is over|\linelabel{alg:rw-rl:wvalidate-5}|
			if marked(cur->next): |\hfill| ## cur is logically deleted?|\linelabel{alg:rw-rl:wvalidate-6}|
				LNode *next = unmark(cur->next)|\linelabel{alg:rw-rl:wvalidate-7}|
				CAS(prev, cur, next) |\hfill|## try to remove it from list|\linelabel{alg:rw-rl:wvalidate-8}|
				cur = next |\hfill|## and continue traversing the list|\linelabel{alg:rw-rl:wvalidate-9}|
			elif cur->end <= lock->start: |\hfill| ## lock succeeds cur |\linelabel{alg:rw-rl:wvalidate-10}|
				prev = &cur->next |\hfill|## continue traversing ...|\linelabel{alg:rw-rl:wvalidate-11}|
				cur = unmark(*prev) |\hfill|## the list.|\linelabel{alg:rw-rl:wvalidate-12}|
			else: |\hfill|## lock overlaps with cur:|\linelabel{alg:rw-rl:wvalidate-13}|
				DeleteNode(lock); |\hfill|## delete the node ...|\linelabel{alg:rw-rl:wvalidate-14}|
				return 1 |\hfill|## and fail validation|\linelabel{alg:rw-rl:wvalidate-15}|
\end{lstlisting}
\end{figure}

%% file: range-locks.bbl

\begin{thebibliography}{36}


\ifx \showCODEN    \undefined \def \showCODEN     #1{\unskip}     \fi
\ifx \showDOI      \undefined \def \showDOI       #1{#1}\fi
\ifx \showISBNx    \undefined \def \showISBNx     #1{\unskip}     \fi
\ifx \showISBNxiii \undefined \def \showISBNxiii  #1{\unskip}     \fi
\ifx \showISSN     \undefined \def \showISSN      #1{\unskip}     \fi
\ifx \showLCCN     \undefined \def \showLCCN      #1{\unskip}     \fi
\ifx \shownote     \undefined \def \shownote      #1{#1}          \fi
\ifx \showarticletitle \undefined \def \showarticletitle #1{#1}   \fi
\ifx \showURL      \undefined \def \showURL       {\relax}        \fi
\providecommand\bibfield[2]{#2}
\providecommand\bibinfo[2]{#2}
\providecommand\natexlab[1]{#1}
\providecommand\showeprint[2][]{arXiv:#2}

\bibitem[\protect\citeauthoryear{Aarestad, Ching, Thiruvathukal, and
  Choudhary}{Aarestad et~al\mbox{.}}{2006}]%
        {Aarestad06}
\bibfield{author}{\bibinfo{person}{P.~M. Aarestad}, \bibinfo{person}{A. Ching},
  \bibinfo{person}{G.~K. Thiruvathukal}, {and} \bibinfo{person}{A.~N.
  Choudhary}.} \bibinfo{year}{2006}\natexlab{}.
\newblock \showarticletitle{Scalable Approaches for Supporting {MPI-IO}
  Atomicity}. In \bibinfo{booktitle}{\emph{Sixth IEEE International Symposium
  on Cluster Computing and the Grid (CCGRID'06)}}, Vol.~\bibinfo{volume}{1}.
  \bibinfo{pages}{35--42}.
\newblock


\bibitem[\protect\citeauthoryear{AT\&T}{AT\&T}{1986}]%
        {unix-manual}
\bibfield{author}{\bibinfo{person}{AT\&T}.} \bibinfo{year}{1986}\natexlab{}.
\newblock \bibinfo{title}{{UNIX System V User's Manual Volume 1}}.
\newblock
\newblock
\urldef\tempurl%
\url{http://bitsavers.trailing-edge.com/pdf/att/3b1/999-801-312IS_ATT_UNIX_PC_System_V_Users_Manual_Volume_1.pdf}
\showURL{%
\tempurl}
\newblock
\shownote{Accessed: 2019-04-15.}


\bibitem[\protect\citeauthoryear{Boyd-Wickizer, Clements, Mao, Pesterev,
  Kaashoek, Morris, and Zeldovich}{Boyd-Wickizer et~al\mbox{.}}{2010}]%
        {BWC10}
\bibfield{author}{\bibinfo{person}{Silas Boyd-Wickizer},
  \bibinfo{person}{Austin~T. Clements}, \bibinfo{person}{Yandong Mao},
  \bibinfo{person}{Aleksey Pesterev}, \bibinfo{person}{M.~Frans Kaashoek},
  \bibinfo{person}{Robert Morris}, {and} \bibinfo{person}{Nickolai Zeldovich}.}
  \bibinfo{year}{2010}\natexlab{}.
\newblock \showarticletitle{An Analysis of {Linux} Scalability to Many Cores}.
  In \bibinfo{booktitle}{\emph{Proceedings of the USENIX Conference on
  Operating Systems Design and Implementation (OSDI)}}. \bibinfo{pages}{1--16}.
\newblock


\bibitem[\protect\citeauthoryear{Bueso}{Bueso}{2017}]%
        {Bue17}
\bibfield{author}{\bibinfo{person}{Davidlohr Bueso}.}
  \bibinfo{year}{2017}\natexlab{}.
\newblock \bibinfo{title}{locking: Introduce range reader/writer lock}.
\newblock \bibinfo{howpublished}{\url{https://lwn.net/Articles/722741/}, May
  15, 2017}.
\newblock
\newblock
\shownote{Accessed: 2018-10-29.}


\bibitem[\protect\citeauthoryear{Bueso}{Bueso}{2018}]%
        {Bue18}
\bibfield{author}{\bibinfo{person}{Davidlohr Bueso}.}
  \bibinfo{year}{2018}\natexlab{}.
\newblock \bibinfo{title}{mm: towards parallel address space operations}.
\newblock \bibinfo{howpublished}{\url{https://lwn.net/Articles/746537/}, Feb 5,
  2018}.
\newblock
\newblock
\shownote{Accessed: 2019-04-15.}


\bibitem[\protect\citeauthoryear{Clements, Kaashoek, and Zeldovich}{Clements
  et~al\mbox{.}}{2012}]%
        {CKZ12}
\bibfield{author}{\bibinfo{person}{Austin~T. Clements},
  \bibinfo{person}{M.~Frans Kaashoek}, {and} \bibinfo{person}{Nickolai
  Zeldovich}.} \bibinfo{year}{2012}\natexlab{}.
\newblock \showarticletitle{Scalable Address Spaces Using {RCU} Balanced
  Trees}. In \bibinfo{booktitle}{\emph{Proceedings of the Seventeenth
  International Conference on Architectural Support for Programming Languages
  and Operating Systems (ASPLOS)}}. \bibinfo{pages}{199--210}.
\newblock


\bibitem[\protect\citeauthoryear{Clements, Kaashoek, and Zeldovich}{Clements
  et~al\mbox{.}}{2013}]%
        {Clements13}
\bibfield{author}{\bibinfo{person}{Austin~T. Clements},
  \bibinfo{person}{M.~Frans Kaashoek}, {and} \bibinfo{person}{Nickolai
  Zeldovich}.} \bibinfo{year}{2013}\natexlab{}.
\newblock \showarticletitle{{RadixVM}: Scalable Address Spaces for
  Multithreaded Applications}. In \bibinfo{booktitle}{\emph{Proceedings of the
  ACM European Conference on Computer Systems (EuroSys)}}.
  \bibinfo{pages}{211--224}.
\newblock


\bibitem[\protect\citeauthoryear{Corbet}{Corbet}{2014}]%
        {linux-locks}
\bibfield{author}{\bibinfo{person}{Jonathan Corbet}.}
  \bibinfo{year}{2014}\natexlab{}.
\newblock \bibinfo{title}{{MCS} locks and qspinlocks}.
\newblock \bibinfo{howpublished}{\url{https://lwn.net/Articles/590243}, March
  11, 2014}.
\newblock
\newblock
\shownote{Accessed: 2018-10-29.}


\bibitem[\protect\citeauthoryear{Corbet}{Corbet}{2017}]%
        {Cor17}
\bibfield{author}{\bibinfo{person}{Jonathan Corbet}.}
  \bibinfo{year}{2017}\natexlab{}.
\newblock \bibinfo{title}{Range reader/writer locks for the kernel}.
\newblock \bibinfo{howpublished}{\url{https://lwn.net/Articles/724502}, June 5,
  2017}.
\newblock
\newblock
\shownote{Accessed: 2018-09-28.}


\bibitem[\protect\citeauthoryear{Dice}{Dice}{2017}]%
        {Dice17}
\bibfield{author}{\bibinfo{person}{Dave Dice}.}
  \bibinfo{year}{2017}\natexlab{}.
\newblock \showarticletitle{Malthusian Locks}. In
  \bibinfo{booktitle}{\emph{Proceedings of the ACM European Conference on
  Computer Systems (EuroSys)}}. \bibinfo{pages}{314--327}.
\newblock


\bibitem[\protect\citeauthoryear{Dice and Kogan}{Dice and Kogan}{2019a}]%
        {DK19b}
\bibfield{author}{\bibinfo{person}{Dave Dice} {and} \bibinfo{person}{Alex
  Kogan}.} \bibinfo{year}{2019}\natexlab{a}.
\newblock \showarticletitle{{BRAVO}: Biased Locking for Reader-writer Locks}.
  In \bibinfo{booktitle}{\emph{Proceedings of the Usenix Annual Technical
  Conference (USENIX ATC)}}. \bibinfo{pages}{315--328}.
\newblock


\bibitem[\protect\citeauthoryear{Dice and Kogan}{Dice and Kogan}{2019b}]%
        {DK19}
\bibfield{author}{\bibinfo{person}{Dave Dice} {and} \bibinfo{person}{Alex
  Kogan}.} \bibinfo{year}{2019}\natexlab{b}.
\newblock \showarticletitle{Compact {NUMA}-aware Locks}. In
  \bibinfo{booktitle}{\emph{Proceedings of the ACM European Conference on
  Computer Systems (EuroSys)}}. \bibinfo{pages}{12:1--12:15}.
\newblock


\bibitem[\protect\citeauthoryear{Dufour}{Dufour}{2017}]%
        {Duf17}
\bibfield{author}{\bibinfo{person}{Laurent Dufour}.}
  \bibinfo{year}{2017}\natexlab{}.
\newblock \bibinfo{title}{Replace mmap\_sem by a range lock}.
\newblock \bibinfo{howpublished}{\url{https://lwn.net/Articles/723648/}, May
  24, 2017}.
\newblock
\newblock
\shownote{Accessed: 2018-10-29.}


\bibitem[\protect\citeauthoryear{et~al.}{et~al.}{2020}]%
        {tor20}
\bibfield{author}{\bibinfo{person}{L.~Torvalds et al.}}
  \bibinfo{year}{2020}\natexlab{}.
\newblock \bibinfo{title}{Linux source code.}
\newblock \bibinfo{howpublished}{\url{http://www.kernel.org/}}.
\newblock
\newblock
\shownote{Accessed: 2020-03-10.}


\bibitem[\protect\citeauthoryear{Faleiro and Abadi}{Faleiro and Abadi}{2017}]%
        {FA17}
\bibfield{author}{\bibinfo{person}{Jose~M. Faleiro} {and}
  \bibinfo{person}{Daniel~J. Abadi}.} \bibinfo{year}{2017}\natexlab{}.
\newblock \showarticletitle{Latch-free Synchronization in Database Systems:
  Silver Bullet or Fool's Gold?}. In \bibinfo{booktitle}{\emph{Proceedings of
  Conference on Innovative Data Systems Research (CIDR)}}.
\newblock


\bibitem[\protect\citeauthoryear{Fraser}{Fraser}{2004}]%
        {Fra04}
\bibfield{author}{\bibinfo{person}{K. Fraser}.}
  \bibinfo{year}{2004}\natexlab{}.
\newblock \emph{\bibinfo{title}{Practical lock-freedom}}.
\newblock \bibinfo{thesistype}{Ph.D. Dissertation}. \bibinfo{school}{University
  of Cambridge}.
\newblock


\bibitem[\protect\citeauthoryear{Graefe}{Graefe}{2007}]%
        {Graefe07}
\bibfield{author}{\bibinfo{person}{Goetz Graefe}.}
  \bibinfo{year}{2007}\natexlab{}.
\newblock \showarticletitle{Hierarchical locking in {B}-tree indexes}. In
  \bibinfo{booktitle}{\emph{BTW}}.
\newblock


\bibitem[\protect\citeauthoryear{Gramoli}{Gramoli}{2015}]%
        {Gra15}
\bibfield{author}{\bibinfo{person}{Vincent Gramoli}.}
  \bibinfo{year}{2015}\natexlab{}.
\newblock \showarticletitle{More Than You Ever Wanted to Know About
  Synchronization: {S}ynchrobench, Measuring the Impact of the Synchronization
  on Concurrent Algorithms}. In \bibinfo{booktitle}{\emph{Proceedings of the
  20th ACM SIGPLAN Symposium on Principles and Practice of Parallel Programming
  (PPoPP)}}.
\newblock


\bibitem[\protect\citeauthoryear{Harris}{Harris}{2001}]%
        {Harris01}
\bibfield{author}{\bibinfo{person}{Timothy~L. Harris}.}
  \bibinfo{year}{2001}\natexlab{}.
\newblock \showarticletitle{A Pragmatic Implementation of Non-blocking
  Linked-Lists}. In \bibinfo{booktitle}{\emph{Proceedings of the 15th
  International Conference on Distributed Computing (DISC)}}.
  \bibinfo{pages}{300--314}.
\newblock


\bibitem[\protect\citeauthoryear{Hart, McKenney, Brown, and Walpole}{Hart
  et~al\mbox{.}}{2007}]%
        {HMBW07}
\bibfield{author}{\bibinfo{person}{Thomas~E. Hart}, \bibinfo{person}{Paul~E.
  McKenney}, \bibinfo{person}{Angela~Demke Brown}, {and}
  \bibinfo{person}{Jonathan Walpole}.} \bibinfo{year}{2007}\natexlab{}.
\newblock \showarticletitle{Performance of Memory Reclamation for Lockless
  Synchronization}.
\newblock \bibinfo{journal}{\emph{J. Parallel Distrib. Comput.}}
  \bibinfo{volume}{67}, \bibinfo{number}{12} (\bibinfo{year}{2007}),
  \bibinfo{pages}{1270--1285}.
\newblock


\bibitem[\protect\citeauthoryear{Herlihy, Lev, Luchangco, and Shavit}{Herlihy
  et~al\mbox{.}}{2007}]%
        {HLL07}
\bibfield{author}{\bibinfo{person}{Maurice Herlihy}, \bibinfo{person}{Yossi
  Lev}, \bibinfo{person}{Victor Luchangco}, {and} \bibinfo{person}{Nir
  Shavit}.} \bibinfo{year}{2007}\natexlab{}.
\newblock \showarticletitle{A Simple Optimistic Skiplist Algorithm}. In
  \bibinfo{booktitle}{\emph{Proceedings of the 14th International Conference on
  Structural Information and Communication Complexity (SIROCCO)}}.
\newblock


\bibitem[\protect\citeauthoryear{Kara}{Kara}{2013}]%
        {Kar13}
\bibfield{author}{\bibinfo{person}{Jan Kara}.} \bibinfo{year}{2013}\natexlab{}.
\newblock \bibinfo{title}{lib: {I}mplement range locks}.
\newblock \bibinfo{howpublished}{\url{https://lkml.org/lkml/2013/1/31/483},
  January 31, 2013}.
\newblock
\newblock
\shownote{Accessed: 2018-09-28.}


\bibitem[\protect\citeauthoryear{Kashyap, Min, and Kim}{Kashyap
  et~al\mbox{.}}{2017}]%
        {KMK17}
\bibfield{author}{\bibinfo{person}{Sanidhya Kashyap}, \bibinfo{person}{Changwoo
  Min}, {and} \bibinfo{person}{Taesoo Kim}.} \bibinfo{year}{2017}\natexlab{}.
\newblock \showarticletitle{Scalable {NUMA}-aware Blocking Synchronization
  Primitives}. In \bibinfo{booktitle}{\emph{Proceedings of the Usenix Annual
  Technical Conference ({USENIX} {ATC})}}.
\newblock


\bibitem[\protect\citeauthoryear{Kim, Kim, Kang, Lee, Park, and Kim}{Kim
  et~al\mbox{.}}{2019}]%
        {KKK19}
\bibfield{author}{\bibinfo{person}{June-Hyung Kim}, \bibinfo{person}{Jangwoong
  Kim}, \bibinfo{person}{Hyeongu Kang}, \bibinfo{person}{Chang-Gyu Lee},
  \bibinfo{person}{Sungyong Park}, {and} \bibinfo{person}{Youngjae Kim}.}
  \bibinfo{year}{2019}\natexlab{}.
\newblock \showarticletitle{{pNOVA}: Optimizing Shared File {I/O} Operations of
  {NVM} File System on Manycore Servers}. In
  \bibinfo{booktitle}{\emph{Proceedings of the 10th ACM SIGOPS Asia-Pacific
  Workshop on Systems (APSys)}}. \bibinfo{pages}{1--7}.
\newblock


\bibitem[\protect\citeauthoryear{Lomet and Mokbel}{Lomet and Mokbel}{2009}]%
        {Lomet09}
\bibfield{author}{\bibinfo{person}{David Lomet} {and}
  \bibinfo{person}{Mohamed~F. Mokbel}.} \bibinfo{year}{2009}\natexlab{}.
\newblock \showarticletitle{Locking Key Ranges with Unbundled Transaction
  Services}.
\newblock \bibinfo{journal}{\emph{Proc. VLDB Endow.}} \bibinfo{volume}{2},
  \bibinfo{number}{1} (\bibinfo{date}{Aug.} \bibinfo{year}{2009}),
  \bibinfo{pages}{265--276}.
\newblock
\showISSN{2150-8097}


\bibitem[\protect\citeauthoryear{Lomet}{Lomet}{1993}]%
        {Lomet93}
\bibfield{author}{\bibinfo{person}{David~B. Lomet}.}
  \bibinfo{year}{1993}\natexlab{}.
\newblock \showarticletitle{Key Range Locking Strategies for Improved
  Concurrency}. In \bibinfo{booktitle}{\emph{Proceedings of the 19th
  International Conference on Very Large Data Bases}}
  \emph{(\bibinfo{series}{VLDB '93})}. \bibinfo{publisher}{Morgan Kaufmann
  Publishers Inc.}, \bibinfo{address}{San Francisco, CA, USA},
  \bibinfo{pages}{655--664}.
\newblock
\showISBNx{1-55860-152-X}


\bibitem[\protect\citeauthoryear{Mao, Morris, and Kaashoek}{Mao
  et~al\mbox{.}}{2010}]%
        {MMK10}
\bibfield{author}{\bibinfo{person}{Yandong Mao}, \bibinfo{person}{Robert
  Morris}, {and} \bibinfo{person}{Frans Kaashoek}.}
  \bibinfo{year}{2010}\natexlab{}.
\newblock \bibinfo{booktitle}{\emph{Optimizing {MapReduce} for Multicore
  Architectures}}.
\newblock \bibinfo{type}{{T}echnical {R}eport}. \bibinfo{institution}{MIT}.
\newblock


\bibitem[\protect\citeauthoryear{Mckenney, Boyd-wickizer, and Walpole}{Mckenney
  et~al\mbox{.}}{2012}]%
        {MBW12}
\bibfield{author}{\bibinfo{person}{Paul~E. Mckenney}, \bibinfo{person}{Silas
  Boyd-wickizer}, {and} \bibinfo{person}{Jonathan Walpole}.}
  \bibinfo{year}{2012}\natexlab{}.
\newblock \bibinfo{booktitle}{\emph{{RCU} usage in the {L}inux kernel: One
  decade later}}.
\newblock \bibinfo{type}{{T}echnical {R}eport}.
\newblock


\bibitem[\protect\citeauthoryear{McKenney and Slingwine}{McKenney and
  Slingwine}{1998}]%
        {MS98}
\bibfield{author}{\bibinfo{person}{Paul~E. McKenney} {and}
  \bibinfo{person}{Jack Slingwine}.} \bibinfo{year}{1998}\natexlab{}.
\newblock \showarticletitle{Read-copy-update: Using Execution History to Solve
  Concurrency Problems}. In \bibinfo{booktitle}{\emph{Parallel and Distributed
  Computing and Systems}}. \bibinfo{pages}{509--518}.
\newblock


\bibitem[\protect\citeauthoryear{Michael}{Michael}{2004}]%
        {Michael04}
\bibfield{author}{\bibinfo{person}{Maged~M. Michael}.}
  \bibinfo{year}{2004}\natexlab{}.
\newblock \showarticletitle{Hazard Pointers: Safe Memory Reclamation for
  Lock-Free Objects}.
\newblock \bibinfo{journal}{\emph{IEEE Trans. Parallel Distrib. Syst.}}
  \bibinfo{volume}{15}, \bibinfo{number}{6} (\bibinfo{year}{2004}),
  \bibinfo{pages}{491--504}.
\newblock


\bibitem[\protect\citeauthoryear{Mohan}{Mohan}{1990}]%
        {Mohan90}
\bibfield{author}{\bibinfo{person}{C. Mohan}.} \bibinfo{year}{1990}\natexlab{}.
\newblock \showarticletitle{{ARIES/KVL}: A Key-value Locking Method for
  Concurrency Control of Multiaction Transactions Operating on {B-tree}
  Indexes}. In \bibinfo{booktitle}{\emph{Proceedings of the Sixteenth
  International Conference on Very Large Databases}}.
  \bibinfo{publisher}{Morgan Kaufmann Publishers Inc.}, \bibinfo{address}{San
  Francisco, CA, USA}, \bibinfo{pages}{392--405}.
\newblock
\showISBNx{0-55860-149-X}


\bibitem[\protect\citeauthoryear{Mohan, Haderle, Lindsay, Pirahesh, and
  Schwarz}{Mohan et~al\mbox{.}}{1992}]%
        {Mohan92}
\bibfield{author}{\bibinfo{person}{C. Mohan}, \bibinfo{person}{Don Haderle},
  \bibinfo{person}{Bruce Lindsay}, \bibinfo{person}{Hamid Pirahesh}, {and}
  \bibinfo{person}{Peter Schwarz}.} \bibinfo{year}{1992}\natexlab{}.
\newblock \showarticletitle{{ARIES}: A Transaction Recovery Method Supporting
  Fine-granularity Locking and Partial Rollbacks Using Write-ahead Logging}.
\newblock \bibinfo{journal}{\emph{ACM Trans. Database Syst.}}
  \bibinfo{volume}{17}, \bibinfo{number}{1} (\bibinfo{date}{March}
  \bibinfo{year}{1992}), \bibinfo{pages}{94--162}.
\newblock
\showISSN{0362-5915}


\bibitem[\protect\citeauthoryear{Quinson and Vernier}{Quinson and
  Vernier}{2009}]%
        {Quinson09}
\bibfield{author}{\bibinfo{person}{M. Quinson} {and} \bibinfo{person}{F.
  Vernier}.} \bibinfo{year}{2009}\natexlab{}.
\newblock \showarticletitle{Byte-Range Asynchronous Locking in Distributed
  Settings}. In \bibinfo{booktitle}{\emph{2009 17th Euromicro International
  Conference on Parallel, Distributed and Network-based Processing}}.
  \bibinfo{pages}{191--195}.
\newblock
\showISSN{1066-6192}


\bibitem[\protect\citeauthoryear{Schmuck and Haskin}{Schmuck and
  Haskin}{2002}]%
        {SH02}
\bibfield{author}{\bibinfo{person}{Frank Schmuck} {and} \bibinfo{person}{Roger
  Haskin}.} \bibinfo{year}{2002}\natexlab{}.
\newblock \showarticletitle{{GPFS}: A Shared-Disk File System for Large
  Computing Clusters}. In \bibinfo{booktitle}{\emph{Proceedings of USENIX
  Conference on File and Storage Technologies (FAST)}}.
\newblock


\bibitem[\protect\citeauthoryear{Song, Shi, Liu, Yang, and Chen}{Song
  et~al\mbox{.}}{2013}]%
        {SSL13}
\bibfield{author}{\bibinfo{person}{Xiang Song}, \bibinfo{person}{Jicheng Shi},
  \bibinfo{person}{Ran Liu}, \bibinfo{person}{Jian Yang}, {and}
  \bibinfo{person}{Haibo Chen}.} \bibinfo{year}{2013}\natexlab{}.
\newblock \showarticletitle{Parallelizing Live Migration of Virtual Machines}.
  In \bibinfo{booktitle}{\emph{Proceedings of the 9th ACM SIGPLAN/SIGOPS
  International Conference on Virtual Execution Environments (VEE)}}.
  \bibinfo{pages}{85--96}.
\newblock


\bibitem[\protect\citeauthoryear{Thakur, Ross, and Latham}{Thakur
  et~al\mbox{.}}{2005}]%
        {Thakur05}
\bibfield{author}{\bibinfo{person}{Rajeev Thakur}, \bibinfo{person}{Robert
  Ross}, {and} \bibinfo{person}{Robert Latham}.}
  \bibinfo{year}{2005}\natexlab{}.
\newblock \showarticletitle{Implementing Byte-Range Locks Using {MPI} One-Sided
  Communication}. In \bibinfo{booktitle}{\emph{Recent Advances in Parallel
  Virtual Machine and Message Passing Interface}},
  \bibfield{editor}{\bibinfo{person}{Beniamino Di~Martino},
  \bibinfo{person}{Dieter Kranzlm{\"u}ller}, {and} \bibinfo{person}{Jack
  Dongarra}} (Eds.). \bibinfo{publisher}{Springer Berlin Heidelberg},
  \bibinfo{address}{Berlin, Heidelberg}, \bibinfo{pages}{119--128}.
\newblock


\end{thebibliography}
